\documentclass[twocolumn,prb,aps]{revtex4}

\usepackage{graphicx}
\usepackage{dcolumn}
\usepackage{amsmath}

\newcommand{\figurewidth}{8.3cm}
\newcommand{\beq}{\begin{equation}}
\newcommand{\eeq}{\end{equation}}
\newcommand{\bea}{\begin{eqnarray}}
\newcommand{\eea}{\end{eqnarray}}

\begin{document}

\title[Short Title]{
Path Integral Monte Carlo and Density Functional Molecular Dynamics Simulations of Hot, Dense Helium}

\author{B. Militzer}
\affiliation{Departments of Earth and Planetary Science and Astronomy, University of California, Berkeley, CA 94720, USA}

\begin{abstract}
Two first-principles simulation techniques, path integral Monte Carlo
(PIMC) and density functional molecular dynamics (DFT-MD), are applied
to study hot, dense helium in the density-temperature range of 0.387
-- 5.35 g$\,$cm$^{-3}$ and 500 K -- 1.28$\times$10$^8$ K. One coherent
equation of state (EOS) is derived by combining DFT-MD data at lower
temperatures with PIMC results at higher temperatures. Good agreement
between both techniques is found in an intermediate temperature
range. For the highest temperatures, the PIMC results converge to the
Debye-H\"uckel limiting law. In order derive the entropy, a
thermodynamically consistent free energy fit is introduced that
reproduces the internal energies and pressure derived from the
first-principles simulations. The equation of state is presented in
form of a table as well as a fit and is compared with chemical
models. In addition, the structure of the fluid is analyzed using pair
correlation functions. Shock Hugoniot curves are compared with recent
laser shock wave experiments.
\end{abstract}

\date{\today }


\maketitle

\section{Introduction}

After hydrogen, helium is the second most common element in the
universe. While it rarely occurs in pure form in nature, it is an
endmember of hydrogen-helium mixtures (HHM) that are the prevalent
component in solar and extrasolar giant gas planets. The
characterization of helium's properties at extreme temperature and
pressure conditions is therefore important to study planetary
interiors and especially relevant for answering the question whether
HHM phase-separate in giant planet
interiors~\cite{Stevenson77a,Stevenson77b}. In most planetary models,
the equation of state (EOS) of HHM was inferred from the linear mixing
approximation at constant pressure and temperature using the EOSs of
pure hydrogen and helium. The latter is the central topic of this
article.

Hydrogen and helium share some common properties. Both are very
light and exhibit rich quantum properties at low temperature. More
importantly for this paper, the helium atom and the deuterium molecule
have similar masses and both have two elemental excitation mechanisms
that determine their behavior at high temperature. The helium atom has
two ionization stages while deuterium molecules can dissociate and the
resulting atoms can be ionized. However, helium is without question
simpler to characterize at high pressure. The crystal structure is
hexagonal closed-packed under most $(P,T)$ conditions~\cite{Ma88,Lo93}
while in solid hydrogen, different degrees of molecular rotational
ordering lead to several phases that deviate from the
h.c.p. structure. Hydrogen is expected to turn metallic at a few
hundred GPa while a much larger bandgap must be closed in helium,
which is predicted to occur above 10$\,$000 GPa~\cite{YMR84,kowalski07}.

Given the relative simplicity of helium's high pressure properties one
expects that there would be less of a controversy in the EOS than for
hydrogen. This makes helium a good material to test novel experimental
and theoretical approaches. For hydrogen, the first laser shock
experiments that reached megabar pressures had predicted that the material
would be highly compressible under shock conditions and reach
densities six times higher than initial state~\cite{Si97,Co98}. Later
experiments~\cite{Kn01,Kn03,Belov2002,BoriskovNellis05} showed reduced
compression ratios close to 4.3, which were in good agreement with
first-principles calculations~\cite{Le97,MC00,Bonev2004}. In the case
of helium, there is very good agreement between the early shock
experiments by Nellis {\em et al.}\cite{nellis84} and first-principles
calculation~\cite{Mi06}. 

Recently the first laser shock experiments were performed on
precompressed helium samples~\cite{eggert08}. The measurements
confirmed the theoretically predicted trend~\cite{Mi06} that the shock
compression ratio is reduced with increasing precompression. However,
there is discrepancy in the magnitude of the compression. Shock
measurements~\cite{eggert08} without precompression showed compression
ratios of about 6 while first-principles simulation~\cite{Mi06}
predicted only 5.24(4). The discrepancy between theoretical and
experimental predictions is reduced for higher precompressions. For
samples that were precompressed to 3.4-fold the ambient density,
theory and experiment are in agreement.

The properties of fluid helium change from hard-sphere liquid at low
pressure and temperature to ultimately a two-component plasma, after
ionization has occurred at high pressure and temperature. The
associated insulator-to-metal transition has been the topic of three
recent theoretical studies that all relied on DFT methods. Kietzmann
{\it et al.}~\cite{Kietzmann07} studied the rise in electrical
conductivity using the Kubo-Greenwood formula and compared with
shock-wave experiments by Ternovoi {\it et
al.}~\cite{Ternovoi04}. Kowalski {\it et al.}~\cite{kowalski07}
studied dense helium in order to characterize the atmosphere of white
dwarfs. This paper went beyond the generalized gradient approximation by
considering hybrid functionals. Stixrude and
Jeanloz~\cite{StixrudeJeanloz08} studied the band gap closure in the
dense fluid helium at over a wide range of densities including
conditions of giant planet interiors.

This article provides the EOS for fluid helium over a wide range of
temperatures (500$\,$K--1.28$\times$10$^8\,$K) and densities
(0.387--5.35 g$\,$cm$^{-3}$ corresponding to a Wigner-Seitz radius
interval of r$_s$=2.4--1.0 where $\frac{4}{3}\pi r_s^3=V/N_e$) by
combining two first-principles simulation methods, path integral Monte
Carlo (PIMC) at higher temperatures with density functional molecular
dynamics (DFT-MD) at lower temperatures. The temperature range was
significantly extended compared to our earlier work~\cite{Mi06} that
focused exclusively on shock properties alone. Here, the region of
validity of both first-principles methods is analyzed and good
agreement for EOS at intermediate temperatures is demonstrated. The
PIMC calculations have been extended to much higher temperatures until
good agreement with the Debye-H\"uckel limiting law is found. In the
density interval under consideration, the entire EOS of
nonrelativistic, fluid helium has been mapped out from first
principles. All EOS data are combined into one thermodynamically
consistent fit for the free energy and the entropy is derived. The
structure of the fluid is analyzed using pair correlation functions
and, finally, additional results for shock Hugoniot curves are
presented.

\begin{figure}[!]
\includegraphics[angle=0,width=\figurewidth]{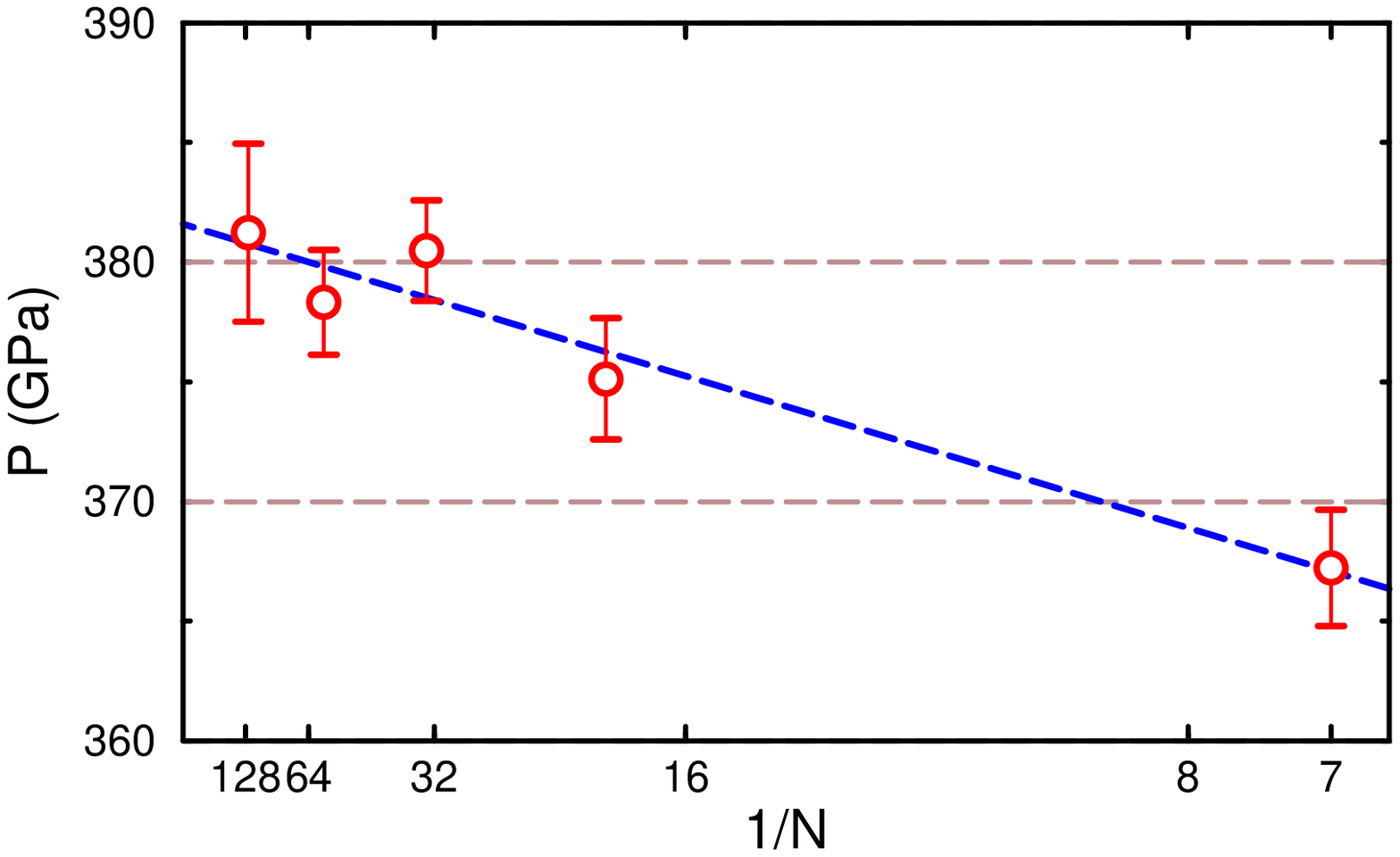}
\includegraphics[angle=0,width=\figurewidth]{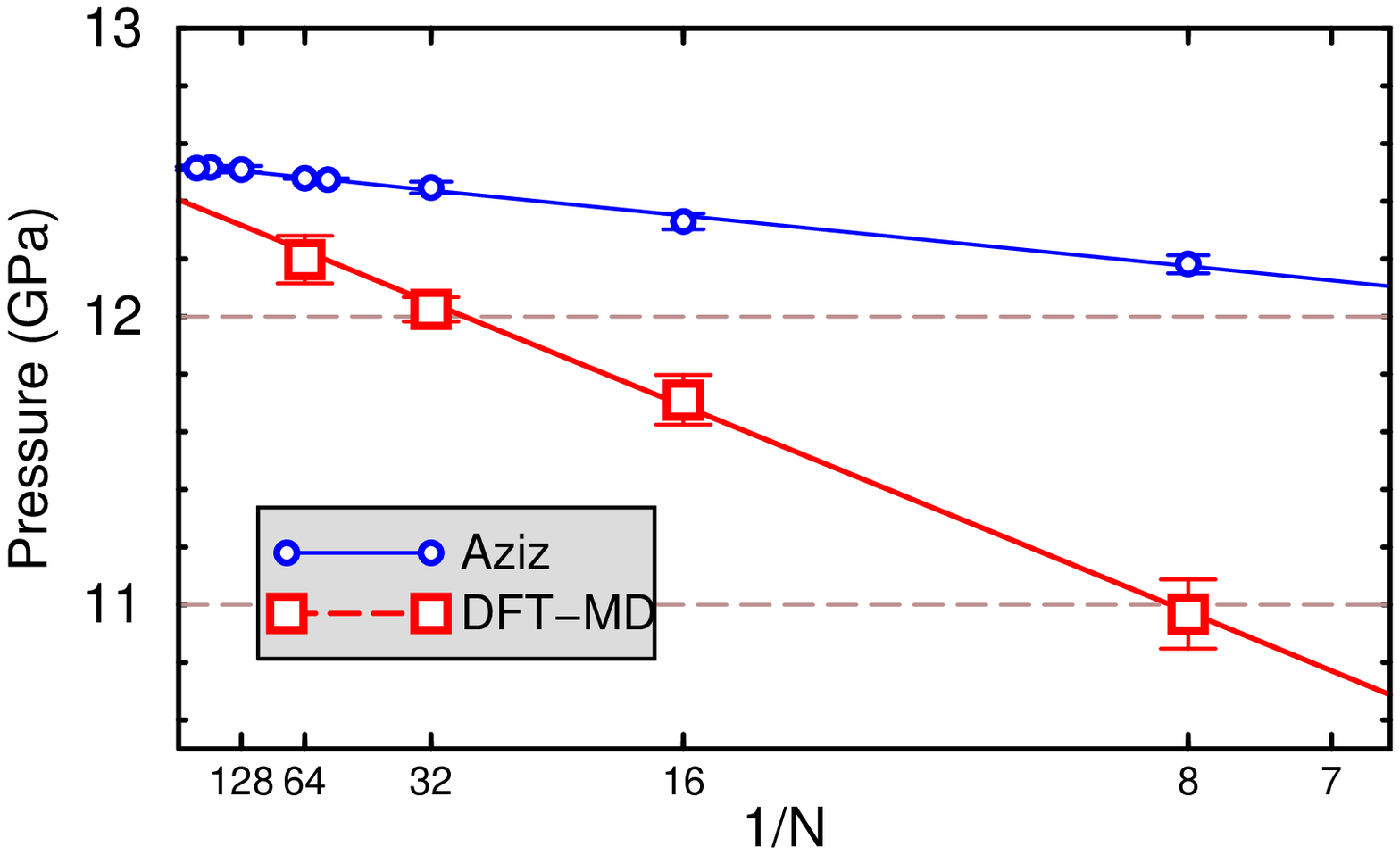}
\caption{ 
       The upper panel shows the finite size dependence of the
       pressure as function of the number atoms, $N$, as predicted
       from PIMC simulations with free-particle nodes at
      T=$125\,000\,$K and $r_s=1.86$. The lower panel compares the
       finite size dependence of DFT-MD simulations and classical
       Monte Carlo calculations using the Aziz pair potential at
       T=10$\,$000$\,$K and $r_s=2.4$.}
\label{fig1}
\end{figure}

\section{Methods}

Path integral Monte Carlo~\cite{Ce95} is the most appropriate and
efficient first-principles simulation techniques for quantum system
with thermal excitations. Electrons and nuclei are treated equally as
paths, although the zero-point motion of the nuclei as well as
exchange effects are negligible for the temperatures under
consideration. The Coulomb interaction between electrons and nuclei is
introduced using pair density matrices that we derived using the
eigenstates of the two-body Coulomb problem~\cite{Po88}. The periodic
images were treated using an optimized Ewald break-up~\cite{Na95} that
we applied to the pair action~\cite{MG06}. The explicit treatment of
electrons as paths leads to the fermion sign problem, which requires
one to introduce the only uncontrolled approximation in this method,
the fixed node approximation~\cite{Ce91,Ce96}. We use the nodes from the
free-particle density matrix and from a variational density
matrix~\cite{MP00}. Besides this approximation, all correlation
effects are included in PIMC, which for example leads an exact
treatment of the isolated helium atom.

The DFT-MD simulations were performed with either the CPMD
code~\cite{CPMD} using local Troullier-Martins norm-conserving
pseudopotentials~\cite{TM91} or with the Vienna ab initio simulation
package~\cite{VASP} using the projector augmented-wave
method~\cite{PAW}. The nuclei were propagated using Born-Oppenheimer
molecular dynamics with forces derived from either the electronic
ground state or by including thermally excited electronic states when
needed. Exchange-correlation effects were described by the
Perdew-Burke-Ernzerhof generalized gradient
approximation~\cite{PBE}. The electronic wavefunctions were expanded
in a plane-wave basis with energy cut-off of 30-50 Hartrees. Most
simulations were performed with $N$=64 using $\Gamma$ point sampling
of the Brillioun zone. An analysis of finite size effect is presented
in the following section.

\section{Equation of state}

An analysis of finite size dependence of the EOS results is important
since all simulation are perform with a finite number of particle in
periodic boundary conditions. Figure~\ref{fig1} gives two examples for
the finite size analysis that we have performed at various $(T,\rho)$
conditions. At $10\,000\,$K and $r_s=2.4$, helium can be characterized
as a hard-sphere fluid. The artificial periodicity of the nuclei
dominate the finite size errors. Simulations with $N=64$ atoms are
sufficiently accurate for the purpose of this study. The DFT-MD
results also agree surprisingly well with classical Monte Carlo
calculation using the Aziz pair potential~\cite{aziz95}, which explains
why both methods give fairly similar Hugoniot curves as long as
electronic excitations are not important~\cite{Mi06}.

\begin{figure}[!]
\includegraphics[angle=0,width=\figurewidth]{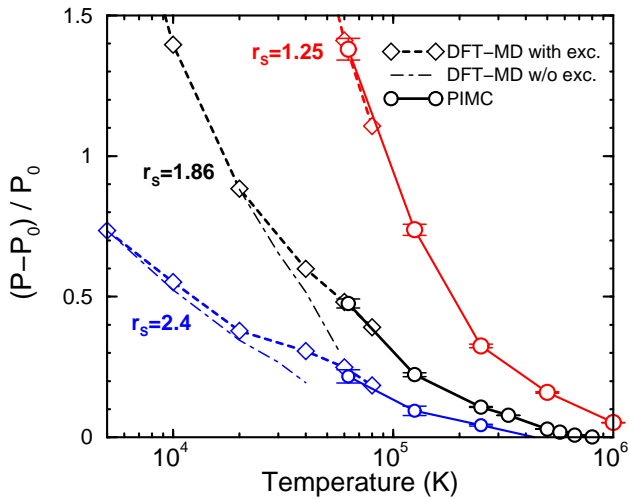}
\caption{ Comparison of the relative excess pressure derived from 
	PIMC (solid lines) and DFT-MD. The dashed and the dash-dotted
	lines show results DFT-MD simulation with and without the
	consideration of thermally excited electronic states,
	respectively. }
\label{PPIMCDFT}
\end{figure}

\begin{figure}[!]
\includegraphics[angle=0,width=\figurewidth]{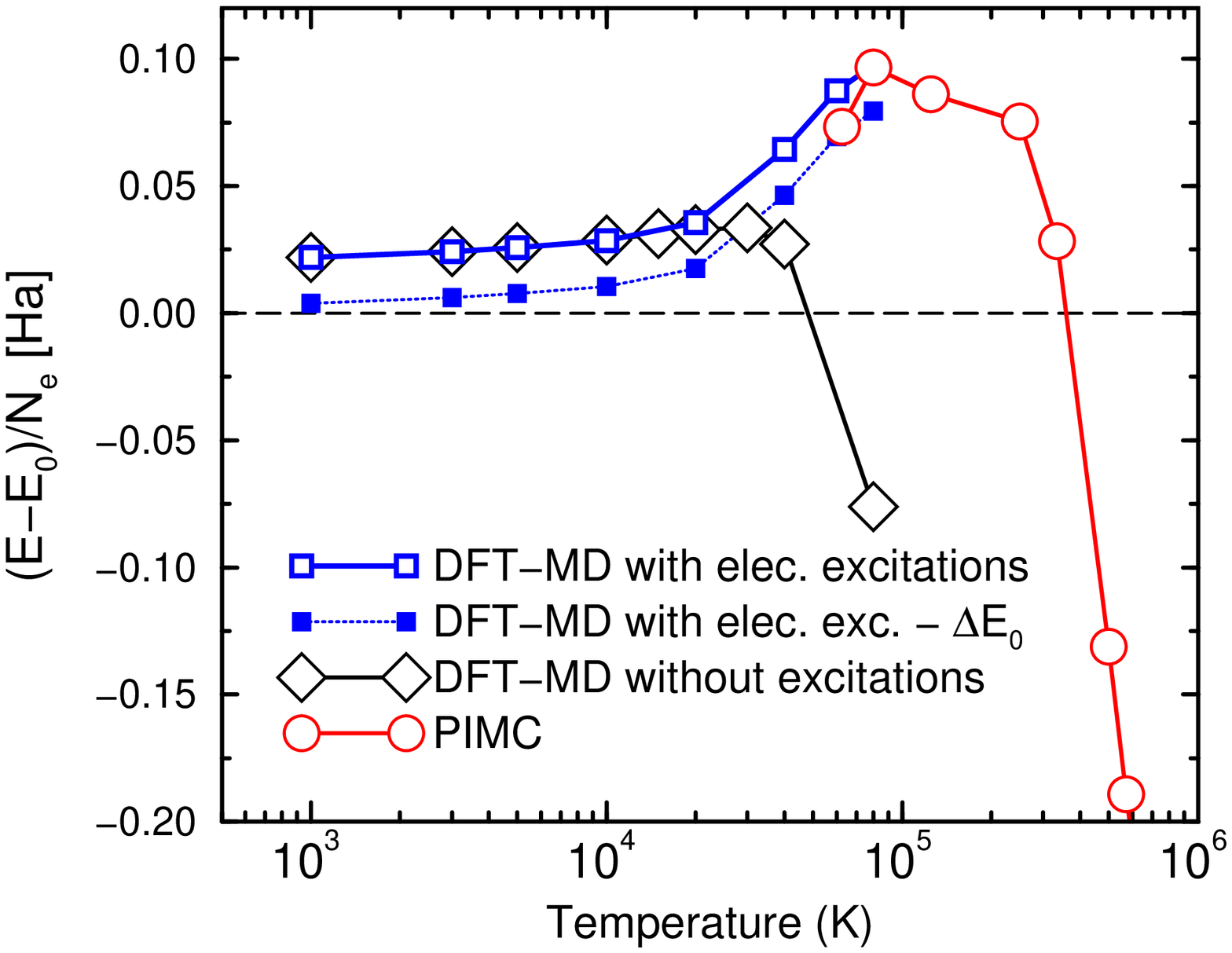}
\includegraphics[angle=0,width=\figurewidth]{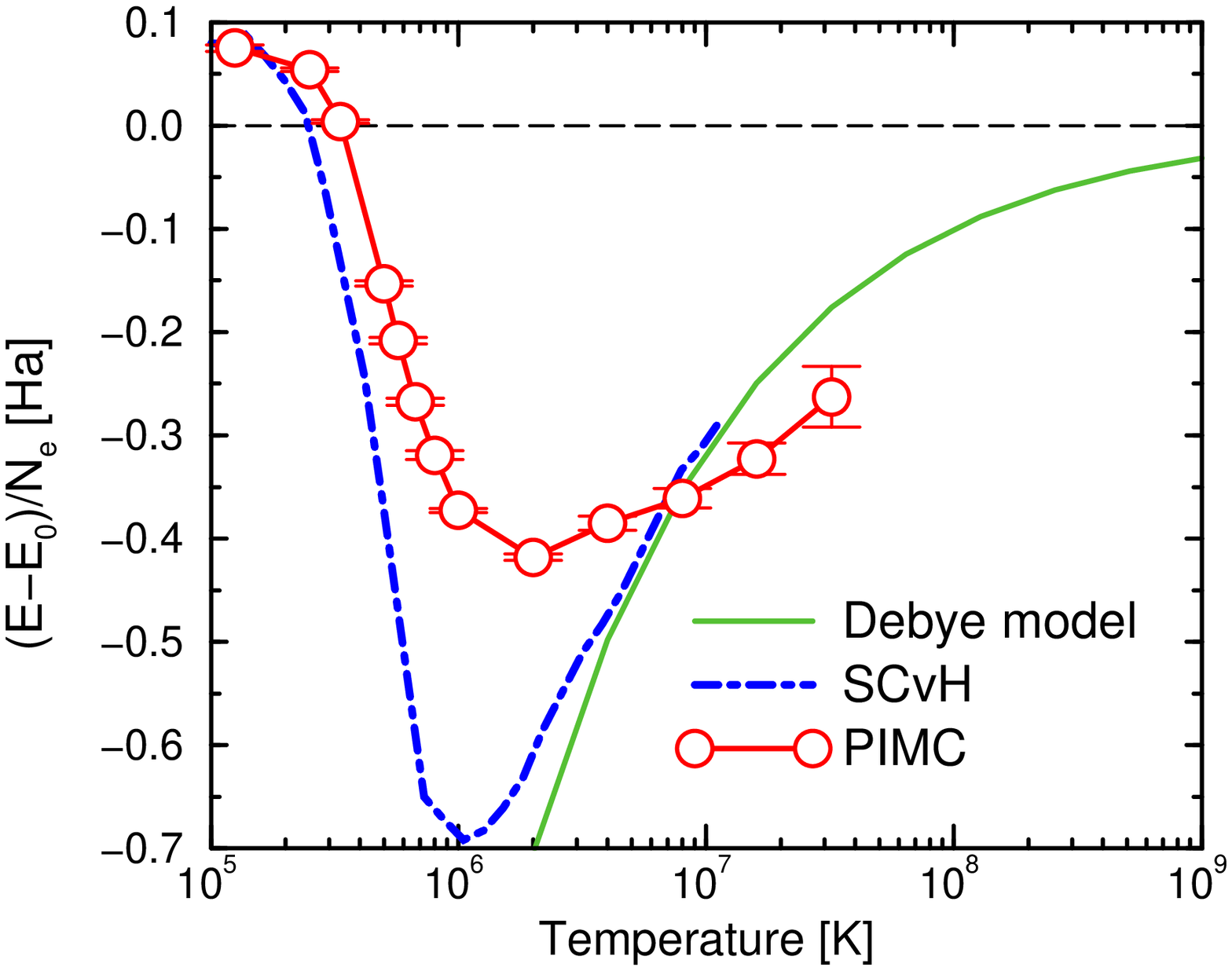}
\caption{ Excess internal energy per electron relative to the ideal 
	plasma model at a density of r$_s$=1.86. The circles show PIMC
	results.  In the upper panel, the open squares and diamonds
	show DFT-MD results with and without thermal population of
	excited electronic states; respectively. The filled squares
	shows DFT-MD results corrected by constant shift corresponding
	to the DFT error of the isolated helium atom. In lower panel,
	PIMC results are compared with the Debye model.  }
\label{EPIMCDFT}
\end{figure}

\begin{table}
\caption{EOS table with pressures, internal and free energies per 
electron derived from (a) DFT-MD with 64 atoms (a uniform $\Delta E/N_e=-0.4909$ Ha correction
was added to account for missing DFT correlation energy in the helium atom),
PIMC with (b) 32 atoms, (c) PIMC with 57 atoms, and (d) Debye-H\"uckel
limiting law. The numbers in brackets indicate the statistical
uncertainties of the DFT-MD and PIMC simulations for the corresponding
number trailing digits.}
\label{EOS}
\begin{tabular}{rrrrr}
$r_s$ & $T ({\rm K})$ & $P\,(\rm{GPa})~$ & $E/N_e\,(\rm{Ha})~$ & $F/N_e\,(\rm{Ha})$\\
\colrule
  2.4$^{a}$ &        500 &      1.420(10) &   -1.449873(7) &    -1.4554 \\
  2.4$^{a}$ &       1000 &      2.045(14) &  -1.448401(10) &   -1.46135 \\
  2.4$^{a}$ &       3000 &        4.69(3) &    -1.44273(3) &   -1.49126 \\
  2.4$^{a}$ &       5000 &        6.98(4) &    -1.43727(3) &   -1.52534 \\
  2.4$^{a}$ &      10000 &       12.49(4) &    -1.42395(5) &   -1.61873 \\
  2.4$^{a}$ &      20000 &       22.19(8) &   -1.39427(12) &   -1.82386 \\
  2.4$^{a}$ &      40000 &      43.37(11) &     -1.2997(2) &   -2.28643 \\
  2.4$^{a}$ &      60000 &      68.27(10) &     -1.1748(2) &   -2.80627 \\
  2.4$^{a}$ &      80000 &      96.93(12) &    -1.02525(7) &   -3.37236 \\
  2.4$^{b}$ &     125000 &       172.3(6) &    -0.6667(17) &   -4.77369 \\
  2.4$^{b}$ &     250000 &       445.7(6) &       0.477(2) &   -9.31702 \\
  2.4$^{b}$ &     333333 &       651.4(9) &       1.237(3) &   -12.6707 \\
  2.4$^{b}$ &     500000 &    1067.7(1.0) &       2.634(3) &    -19.922 \\
  2.4$^{b}$ &     571428 &      1249.9(9) &       3.216(3) &   -23.1952 \\
  2.4$^{b}$ &     666667 &      1484.2(5) &     3.9507(16) &   -27.6612 \\
  2.4$^{b}$ &     800000 &    1815.5(1.2) &       4.972(4) &   -34.0708 \\
  2.4$^{b}$ &    1$\times10^6$ &      2308.4(7) &       6.470(2) &   -43.9954 \\
  2.4$^{b}$ &    2$\times10^6$ &      4745.2(8) &      13.754(2) &   -97.4249 \\
  2.4$^{b}$ &    4$\times10^6$ &    9587.6(1.2) &      28.102(4) &   -213.949 \\
  2.4$^{d}$ &    8$\times10^6$ &          19253 &          56.72 &   -466.803 \\
  2.4$^{d}$ &   16$\times10^6$ &          38577 &         113.80 &   -1013.36 \\
  2.4$^{d}$ &   32$\times10^6$ &          77205 &         227.87 &   -2184.48 \\
  2.4$^{d}$ &   64$\times10^6$ &         154445 &         455.92 &   -4683.43 \\
  2.4$^{d}$ &  128$\times10^6$ &         308916 &         911.97 &   -10002.6 \\
  2.4$^{d}$ &  256$\times10^6$ &         617849 &        1824.04 &   -21267.5 \\
  2.4$^{d}$ &  512$\times10^6$ &        1235711 &        3648.14 &   -45052.4 \\
  2.4$^{d}$ & 1024$\times10^6$ &        2471430 &        7296.32 &   -95193.5 \\
  2.4$^{d}$ & 2048$\times10^6$ &        4942866 &       14592.67 &    -200489 \\
\end{tabular}
\end{table}

\begin{table}
\caption{Table~\ref{EOS} continued.}
\label{EOS2}
\begin{tabular}{rrrrr}
$r_s$ & $T ({\rm K})$ & $P\,(\rm{GPa})~$ & $E/N_e\,(\rm{Ha})~$ & $F/N_e\,(\rm{Ha})$\\
\colrule
  2.2$^{a}$ &        500 &        2.74(2) &  -1.449495(13) &   -1.45451 \\
  2.2$^{a}$ &       1000 &        3.77(3) &    -1.44787(2) &    -1.4601 \\
  2.2$^{a}$ &       3000 &        7.59(3) &    -1.44186(2) &   -1.48855 \\
  2.2$^{a}$ &       5000 &       10.81(6) &    -1.43615(3) &    -1.5214 \\
  2.2$^{a}$ &      10000 &       18.23(7) &    -1.42256(5) &   -1.61205 \\
  2.2$^{a}$ &      20000 &       31.39(7) &    -1.39256(9) &     -1.812 \\
  2.2$^{a}$ &      40000 &      59.54(13) &   -1.30036(19) &    -2.2635 \\
  2.2$^{a}$ &      60000 &      92.10(11) &   -1.17937(16) &   -2.77064 \\
  2.2$^{a}$ &      80000 &     129.52(10) &   -1.03399(12) &   -3.32222 \\
  2.2$^{b}$ &     125000 &       223.9(7) &    -0.6962(16) &   -4.68439 \\
  2.2$^{b}$ &     250000 &       569.6(7) &     0.3971(16) &   -9.08923 \\
  2.2$^{b}$ &     500000 &      1371.3(6) &     2.5403(14) &   -19.3777 \\
  2.2$^{b}$ &    1$\times10^6$ &      2981.1(7) &     6.3896(17) &    -42.807 \\
  2.2$^{b}$ &    2$\times10^6$ &      6148.3(8) &      13.694(2) &        -95 \\
  2.2$^{b}$ &    4$\times10^6$ &   12438.6(1.7) &      28.059(4) &   -209.039 \\
  2.2$^{d}$ &    8$\times10^6$ &          24982 &          56.67 &   -456.924 \\
  2.2$^{d}$ &   16$\times10^6$ &          50075 &         113.77 &   -993.605 \\
  2.2$^{d}$ &   32$\times10^6$ &         100227 &         227.85 &   -2144.93 \\
  2.2$^{d}$ &   64$\times10^6$ &         200508 &         455.91 &   -4604.15 \\
  2.2$^{d}$ &  128$\times10^6$ &         401053 &         911.96 &   -9843.99 \\
  2.2$^{d}$ &  256$\times10^6$ &         802134 &        1824.03 &   -20951.2 \\
  2.2$^{d}$ &  512$\times10^6$ &        1604287 &        3648.13 &   -44420.5 \\
  2.2$^{d}$ & 1024$\times10^6$ &        3208587 &        7296.31 &   -93928.4 \\
  2.2$^{d}$ & 2048$\times10^6$ &        6417184 &       14592.66 &    -197960 \\
\colrule
    2$^{a}$ &        500 &      6.101(13) &   -1.448584(6) &   -1.45297 \\
    2$^{a}$ &       1000 &        7.59(2) &  -1.446835(11) &   -1.45806 \\
    2$^{a}$ &       3000 &       13.57(5) &    -1.44022(3) &   -1.48466 \\
    2$^{a}$ &       5000 &       18.07(7) &    -1.43433(4) &   -1.51606 \\
    2$^{a}$ &      10000 &      28.62(11) &    -1.42013(8) &   -1.60363 \\
    2$^{a}$ &      20000 &      47.02(12) &   -1.38968(11) &   -1.79771 \\
    2$^{a}$ &      40000 &      84.72(12) &   -1.30039(15) &   -2.23673 \\
    2$^{a}$ &      60000 &     128.66(14) &   -1.18357(12) &   -2.72986 \\
    2$^{a}$ &      80000 &     178.84(19) &     -1.0433(2) &   -3.26574 \\
    2$^{b}$ &     125000 &       297.4(7) &    -0.7291(12) &   -4.58614 \\
    2$^{b}$ &     250000 &       745.9(7) &     0.3112(13) &   -8.84403 \\
    2$^{b}$ &     500000 &      1800.9(8) &     2.4269(14) &   -18.7895 \\
    2$^{b}$ &    1$\times10^6$ &    3941.2(1.0) &     6.2898(17) &   -41.5111 \\
    2$^{b}$ &    2$\times10^6$ &    8163.1(1.5) &      13.621(3) &   -92.3417 \\
    2$^{b}$ &    4$\times10^6$ &       16544(2) &      28.009(4) &   -203.653 \\
    2$^{d}$ &    8$\times10^6$ &          33228 &          56.61 &   -446.089 \\
    2$^{d}$ &   16$\times10^6$ &          66633 &         113.73 &   -971.926 \\
    2$^{d}$ &   32$\times10^6$ &         133390 &         227.82 &   -2101.53 \\
    2$^{d}$ &   64$\times10^6$ &         266868 &         455.88 &   -4517.11 \\
    2$^{d}$ &  128$\times10^6$ &         533797 &         911.95 &    -9669.7 \\
    2$^{d}$ &  256$\times10^6$ &        1067636 &        1824.02 &   -20604.1 \\
    2$^{d}$ &  512$\times10^6$ &        2135303 &        3648.12 &   -43727.2 \\
    2$^{d}$ & 1024$\times10^6$ &        4270628 &        7296.31 &   -92540.1 \\
    2$^{d}$ & 2048$\times10^6$ &        8541271 &       14592.66 &    -195186 \\
\end{tabular}
\end{table}

\begin{table}
\caption{Table~\ref{EOS2} continued.}
\label{EOS3}
\begin{tabular}{rrrrr}
$r_s$ & $T ({\rm K})$ & $P\,(\rm{GPa})~$ & $E/N_e\,(\rm{Ha})~$ & $F/N_e\,(\rm{Ha})$\\
\colrule
 1.86$^{a}$ &       1000 &       13.55(3) &  -1.445347(12) &   -1.45574 \\
 1.86$^{a}$ &       3000 &       21.50(8) &    -1.43837(4) &   -1.48081 \\
 1.86$^{a}$ &       5000 &       28.04(9) &    -1.43196(4) &   -1.51097 \\
 1.86$^{a}$ &      10000 &      41.40(10) &    -1.41740(6) &   -1.59608 \\
 1.86$^{a}$ &      20000 &      65.16(13) &    -1.38638(9) &    -1.7856 \\
 1.86$^{a}$ &      40000 &     112.98(18) &   -1.29844(18) &    -2.2149 \\
 1.86$^{a}$ &      60000 &       167.7(2) &     -1.1854(2) &   -2.69739 \\
 1.86$^{a}$ &      80000 &     229.42(15) &   -1.04980(12) &   -3.22147 \\
 1.86$^{c}$ &     125000 &         378(2) &      -0.743(3) &   -4.51072 \\
 1.86$^{b}$ &     250000 &     918.6(1.3) &     0.2491(18) &   -8.65932 \\
 1.86$^{b}$ &     333333 &    1340.7(1.3) &     0.9607(18) &   -11.7193 \\
 1.86$^{b}$ &     500000 &    2214.5(1.7) &       2.336(2) &   -18.3468 \\
 1.86$^{b}$ &     571428 &        2595(2) &       2.910(3) &   -21.3458 \\
 1.86$^{b}$ &     666667 &        3104(3) &       3.661(4) &   -25.4472 \\
 1.86$^{b}$ &     800000 &        3818(3) &       4.699(4) &   -31.3521 \\
 1.86$^{b}$ &    1$\times10^6$ &    4876.7(1.5) &       6.212(2) &   -40.5306 \\
 1.86$^{b}$ &    2$\times10^6$ &       10128(3) &      13.559(3) &   -90.3193 \\
 1.86$^{b}$ &    4$\times10^6$ &       20550(5) &      27.959(7) &   -199.554 \\
 1.86$^{b}$ &    8$\times10^6$ &       41316(7) &      56.543(9) &   -437.845 \\
 1.86$^{d}$ &   16$\times10^6$ &          82822 &         113.69 &   -955.409 \\
 1.86$^{d}$ &   32$\times10^6$ &         165822 &         227.79 &   -2068.45 \\
 1.86$^{d}$ &   64$\times10^6$ &         331768 &         455.87 &   -4450.87 \\
 1.86$^{d}$ &  128$\times10^6$ &         663625 &         911.93 &    -9537.1 \\
 1.86$^{d}$ &  256$\times10^6$ &        1327312 &        1824.01 &   -20338.8 \\
 1.86$^{d}$ &  512$\times10^6$ &        2654668 &        3648.12 &   -43196.4 \\
 1.86$^{d}$ & 1024$\times10^6$ &        5309366 &        7296.30 &   -91478.5 \\
 1.86$^{d}$ & 2048$\times10^6$ &       10618754 &       14592.66 &    -193062 \\
\colrule
 1.75$^{a}$ &       1000 &       22.14(4) &  -1.443419(13) &   -1.45302 \\
 1.75$^{a}$ &       3000 &      32.35(11) &    -1.43604(4) &   -1.47673 \\
 1.75$^{a}$ &       5000 &      40.66(13) &    -1.42933(5) &   -1.50576 \\
 1.75$^{a}$ &      10000 &      57.60(13) &    -1.41415(6) &   -1.58867 \\
 1.75$^{a}$ &      20000 &      87.06(17) &   -1.38263(11) &   -1.77422 \\
 1.75$^{a}$ &      40000 &       144.6(3) &     -1.2961(3) &   -2.19531 \\
 1.75$^{a}$ &      60000 &     210.43(19) &   -1.18602(16) &   -2.66885 \\
 1.75$^{a}$ &      80000 &       284.6(3) &   -1.05398(16) &   -3.18308 \\
 1.75$^{b}$ &     125000 &     454.4(1.0) &    -0.7647(12) &   -4.44653 \\
 1.75$^{b}$ &     250000 &    1098.0(1.1) &     0.2015(13) &   -8.50508 \\
 1.75$^{b}$ &     500000 &    2639.1(1.6) &     2.2626(17) &   -17.9781 \\
 1.75$^{b}$ &    1$\times10^6$ &    5831.7(2.0) &       6.143(2) &   -39.7109 \\
 1.75$^{b}$ &    2$\times10^6$ &       12139(2) &      13.503(3) &   -88.6228 \\
 1.75$^{b}$ &    4$\times10^6$ &       24656(3) &      27.915(4) &   -196.115 \\
 1.75$^{b}$ &    8$\times10^6$ &       49587(8) &     56.501(10) &   -430.926 \\
 1.75$^{d}$ &   16$\times10^6$ &          99420 &         113.65 &   -941.531 \\
 1.75$^{d}$ &   32$\times10^6$ &         199083 &         227.76 &   -2040.64 \\
 1.75$^{d}$ &   64$\times10^6$ &         398335 &         455.85 &   -4395.24 \\
 1.75$^{d}$ &  128$\times10^6$ &         796789 &         911.92 &   -9425.79 \\
 1.75$^{d}$ &  256$\times10^6$ &        1593662 &        1824.00 &   -20115.4 \\
 1.75$^{d}$ &  512$\times10^6$ &        3187384 &        3648.11 &     -42749 \\
 1.75$^{d}$ & 1024$\times10^6$ &        6374809 &        7296.30 &   -90584.7 \\
 1.75$^{d}$ & 2048$\times10^6$ &       12749648 &       14592.65 &    -191274 \\
\end{tabular}
\end{table}

\begin{table}
\caption{Table~\ref{EOS3} continued.}
\label{EOS4}
\begin{tabular}{rrrrr}
$r_s$ & $T ({\rm K})$ & $P\,(\rm{GPa})~$ & $E/N_e\,(\rm{Ha})~$ & $F/N_e\,(\rm{Ha})$\\
\colrule
  1.5$^{a}$ &       1000 &      73.92(10) &    -1.43440(3) &   -1.44135 \\
  1.5$^{a}$ &       2000 &      84.42(12) &    -1.42977(4) &   -1.45018 \\
  1.5$^{a}$ &       3000 &      92.66(12) &    -1.42576(4) &   -1.46128 \\
  1.5$^{a}$ &       5000 &       107.7(2) &    -1.41813(7) &   -1.48725 \\
  1.5$^{a}$ &      10000 &       137.9(3) &   -1.40115(11) &   -1.56358 \\
  1.5$^{a}$ &      20000 &       189.4(4) &   -1.36731(18) &   -1.73814 \\
  1.5$^{a}$ &      40000 &       285.0(4) &     -1.2819(2) &   -2.13878 \\
  1.5$^{a}$ &      60000 &       387.4(4) &     -1.1802(2) &   -2.58963 \\
  1.5$^{a}$ &      80000 &       505.3(3) &   -1.05690(13) &   -3.07884 \\
  1.5$^{b}$ &     125000 &         770(2) &    -0.7880(16) &   -4.27821 \\
  1.5$^{b}$ &     250000 &    1737.1(1.6) &     0.0907(12) &   -8.11696 \\
  1.5$^{b}$ &     500000 &    4114.7(1.6) &     2.0709(12) &   -17.0558 \\
  1.5$^{b}$ &    1$\times10^6$ &    9147.6(1.6) &     5.9425(12) &   -37.6523 \\
  1.5$^{b}$ &    2$\times10^6$ &       19175(2) &    13.3390(15) &   -84.3488 \\
  1.5$^{b}$ &    4$\times10^6$ &       39062(5) &      27.782(4) &   -187.433 \\
  1.5$^{b}$ &    8$\times10^6$ &      78690(10) &      56.411(8) &   -413.442 \\
  1.5$^{c}$ &   16$\times10^6$ &      157860(7) &     113.544(5) &   -906.451 \\
  1.5$^{d}$ &   32$\times10^6$ &         316058 &         227.68 &   -1970.31 \\
  1.5$^{d}$ &   64$\times10^6$ &         632486 &         455.79 &   -4254.46 \\
  1.5$^{d}$ &  128$\times10^6$ &        1265232 &         911.88 &   -9143.99 \\
  1.5$^{d}$ &  256$\times10^6$ &        2530649 &        1823.97 &   -19550.3 \\
  1.5$^{d}$ &  512$\times10^6$ &        5061428 &        3648.09 &   -41618.1 \\
  1.5$^{d}$ & 1024$\times10^6$ &       10122947 &        7296.28 &   -88324.9 \\
  1.5$^{d}$ & 2048$\times10^6$ &       20245959 &       14592.64 &    -186751 \\
\colrule
 1.25$^{a}$ &       3000 &       331.6(3) &    -1.39652(6) &   -1.42554 \\
 1.25$^{a}$ &       5000 &       360.1(5) &   -1.38761(12) &   -1.44742 \\
 1.25$^{a}$ &      10000 &       418.6(4) &    -1.36798(8) &   -1.51449 \\
 1.25$^{a}$ &      20000 &       515.4(8) &     -1.3306(3) &   -1.67468 \\
 1.25$^{a}$ &      40000 &       683.9(5) &   -1.24615(18) &   -2.05136 \\
 1.25$^{a}$ &      60000 &       865.2(7) &     -1.1504(2) &   -2.47506 \\
 1.25$^{a}$ &      80000 &    1063.2(1.0) &     -1.0378(3) &   -2.93438 \\
 1.25$^{a}$ &     125000 &    1565.4(1.5) &     -0.7817(4) &   -4.05965 \\
 1.25$^{b}$ &     250000 &        3074(3) &    -0.0069(11) &   -7.65192 \\
 1.25$^{b}$ &     500000 &        6999(4) &     1.8502(15) &   -15.9783 \\
 1.25$^{b}$ &    1$\times10^6$ &       15578(3) &     5.6796(11) &   -35.2469 \\
 1.25$^{b}$ &    2$\times10^6$ &       32897(7) &      13.107(3) &   -79.3259 \\
 1.25$^{b}$ &    4$\times10^6$ &      67243(80) &       27.58(3) &   -177.186 \\
 1.25$^{b}$ &    8$\times10^6$ &     135808(19) &      56.263(8) &   -392.772 \\
 1.25$^{b}$ &   16$\times10^6$ &     272614(30) &    113.381(13) &   -864.979 \\
 1.25$^{d}$ &   32$\times10^6$ &         545908 &         227.54 &   -1887.16 \\
 1.25$^{d}$ &   64$\times10^6$ &        1092767 &         455.69 &   -4087.74 \\
 1.25$^{d}$ &  128$\times10^6$ &        2186203 &         911.81 &   -8810.14 \\
 1.25$^{d}$ &  256$\times10^6$ &        4372877 &        1823.92 &   -18883.5 \\
 1.25$^{d}$ &  512$\times10^6$ &        8746088 &        3648.06 &     -40286 \\
 1.25$^{d}$ & 1024$\times10^6$ &       17492411 &        7296.26 &   -85659.8 \\
 1.25$^{d}$ & 2048$\times10^6$ &       34984988 &       14592.63 &    -181419 \\
\end{tabular}
\end{table}

\begin{table}
\caption{Table~\ref{EOS4} continued.}
\label{EOS5}
\begin{tabular}{rrrrr}
$r_s$ & $T ({\rm K})$ & $P\,(\rm{GPa})~$ & $E/N_e\,(\rm{Ha})~$ & $F/N_e\,(\rm{Ha})$\\
\colrule
    1$^{a}$ &       5000 &      1560.1(5) &    -1.29739(8) &   -1.34504 \\
    1$^{a}$ &      10000 &      1681.8(7) &   -1.27401(12) &   -1.40096 \\
    1$^{a}$ &      20000 &    1878.6(1.0) &     -1.2313(2) &    -1.5439 \\
    1$^{a}$ &      40000 &    2217.0(1.8) &     -1.1449(3) &   -1.88916 \\
    1$^{a}$ &      62500 &    2608.7(1.8) &     -1.0456(3) &   -2.33028 \\
    1$^{a}$ &      80000 &        2941(2) &     -0.9554(4) &    -2.7022 \\
    1$^{a}$ &     125000 &        3890(2) &     -0.7276(4) &   -3.73996 \\
    1$^{b}$ &     250000 &        6640(6) &    -0.0380(12) &   -7.04803 \\
    1$^{b}$ &     333333 &        8780(6) &     0.4717(12) &   -9.46272 \\
    1$^{b}$ &     500000 &       13687(5) &     1.6229(11) &   -14.6687 \\
    1$^{b}$ &     571428 &       15920(5) &     2.1410(11) &   -17.0286 \\
    1$^{b}$ &     666667 &       18969(5) &     2.8429(11) &   -20.2721 \\
    1$^{b}$ &     800000 &       23334(5) &     3.8385(12) &   -24.9776 \\
    1$^{b}$ &    1$\times10^6$ &       29972(5) &     5.3367(12) &   -32.3609 \\
    1$^{b}$ &    2$\times10^6$ &      63676(11) &      12.775(2) &   -73.2177 \\
    1$^{b}$ &    4$\times10^6$ &     130941(12) &      27.326(3) &   -164.669 \\
    1$^{b}$ &    8$\times10^6$ &     264847(50) &     56.052(10) &    -367.49 \\
    1$^{c}$ &   16$\times10^6$ &     532209(20) &     113.270(5) &    -814.18 \\
    1$^{c}$ &   32$\times10^6$ &    1066140(60) &    227.350(14) &   -1785.39 \\
    1$^{d}$ &   64$\times10^6$ &        2133644 &         455.50 &   -3884.05 \\
    1$^{d}$ &  128$\times10^6$ &        4269457 &         911.68 &   -8402.62 \\
    1$^{d}$ &  256$\times10^6$ &        8540444 &        1823.83 &   -18068.8 \\
    1$^{d}$ &  512$\times10^6$ &       17081968 &        3647.99 &   -38656.7 \\
    1$^{d}$ & 1024$\times10^6$ &       34164699 &        7296.21 &   -82400.7 \\
    1$^{d}$ & 2048$\times10^6$ &       68329938 &       14592.59 &    -174902 \\
\colrule
\end{tabular}
\end{table}

The upper panel of Fig.~\ref{fig1} shows PIMC results for
$125\,000\,$K where a substantial part of the pressure comes from
excited electrons. They are still coupled to the motion of the nuclei,
which leads to effective screening. In consequence, the finite size
dependence of the pressure is reduced significantly, and a simulation
with $N=16$ atoms exhibit a finite size error of only 1\% compared
with 3\% at lower temperature. This is reason why PIMC simulations
with $16$ atoms already gave a fairly accurate shock Hugoniot
curve~\cite{Mi06}. However, most PIMC results reported in
Tab.~\ref{EOS} were obtained with 32 atoms and some with 57
atoms. Already 32 atoms lead to well converged pressures unless one is
interested at very high temperature above 10$^7$K where all atoms are
ionized and the coupling is very weak. Although the total pressure is
dominated by the kinetic term, the excess pressure shows an increased
finite size dependence that requires simulation with 57 atoms in some
cases. In general, the weak-coupling limit is difficult to study with
finite-size simulations~\cite{MP05}. Also at very high density beyond
the range considered here, electrons approach the limit of an ideal
Fermi gas and form a rigid background. The remaining Coulombic
subsystem of ions is expected to require simulations with several
hundreds of particles~\cite{StringfellowDeWittSlattery90}. In this
regard, the electronic screening makes our simulations affordable.

Figure~\ref{PPIMCDFT} compares the pressures obtained from PIMC and
DFT-MD simulations for several density. Above 20$\,$000 K, excited
electronic state become important. Both first-principles method are in
very good agreement, which is foundation for the coherent EOS reported
in this paper. Reasonably good agreement between PIMC and DFT-MD was
reported for hydrogen earlier~\cite{MC01}. Figure~\ref{PPIMCDFT} is a
stringent test because it compares only the pressure contributions
that result from the particle interactions. When one removes the ideal
gas contributions, $P_0$, has one a bit of a choice for the
corresponding noninteracting system. At very high temperature, one
wants to compare with an ideal Fermi gas of electrons and nuclei. At
low temperature, however, one prefers comparing with an gas of
noninteracting atoms. To combine these to limiting cases, we construct
a simple chemical model that includes neutral atoms, He$^+$ and
He$^{2+}$ ions as well as free electrons. The ionization state is
derived from the Saha equilibrium using the following binding
energies, $E_{\rm He}=-79.0\,$eV, $E_{\rm He+}=-54.4\,$eV. Besides the
binding energies, no other interaction are considered.

This approach smoothly connects the low- and high-temperature
limits. However for the correct interpretation of the presented
graphs, it should be pointed out that the pressures and energies of
the ideal model depend on the Saha ionization equilibrium. If the
ideal system exhibits a higher degree of ionization than the
simulation results, then this alone can lead to negative excess pressures
and energies, which one would normally attribute exclusively to the
interaction of free electrons and ions. This effect becomes clear in
Fig.~\ref{EPIMCDFT} where even the DFT-MD results without excited
electrons exhibit a negative excess internal energy at 80$\,$000 K.

Figure~\ref{EPIMCDFT} exhibits the missing correlation energies in DFT
GGA, which underestimates the binding energy in the atom by $\Delta
E_0$=0.98 eV. This is the main reason for the disagreement with the
ideal plasma model in the low-temperature limit (1.20 eV/atom), while
internal energy increase due to the compression to $r_s=1.86$ amounts
only to 0.22 eV/atom. To correct for missing correlation energy, we
applied a uniform correction of $-\Delta E_0$ to all DFT results
discussed later. We may assume that the correction to DFT depends only
weakly on temperature and density. To determine its precise amount
more accurately is difficult and goes beyond the scope of this
article.

Despite this DFT insufficiency, one finds reasonably good agreement in
the internal energies reported by PIMC and
DFT-MD. Figure~\ref{EPIMCDFT} shows that both methods report very
similar increases in the energy resulting from thermal excitations of
electrons, which is the basis for constructing one EOS table.

\begin{figure}[!]
\includegraphics[angle=0,width=\figurewidth]{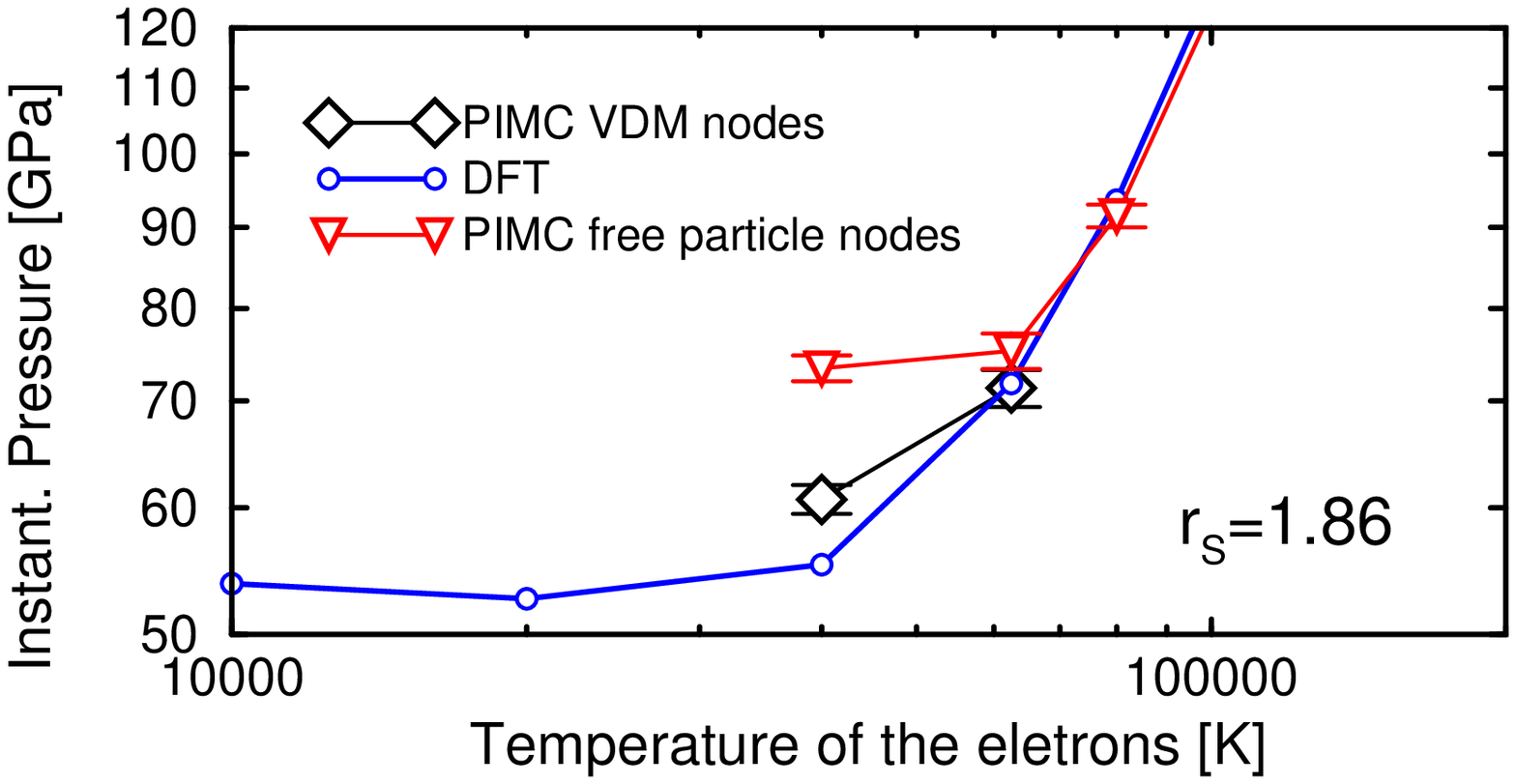}
\includegraphics[angle=0,width=\figurewidth]{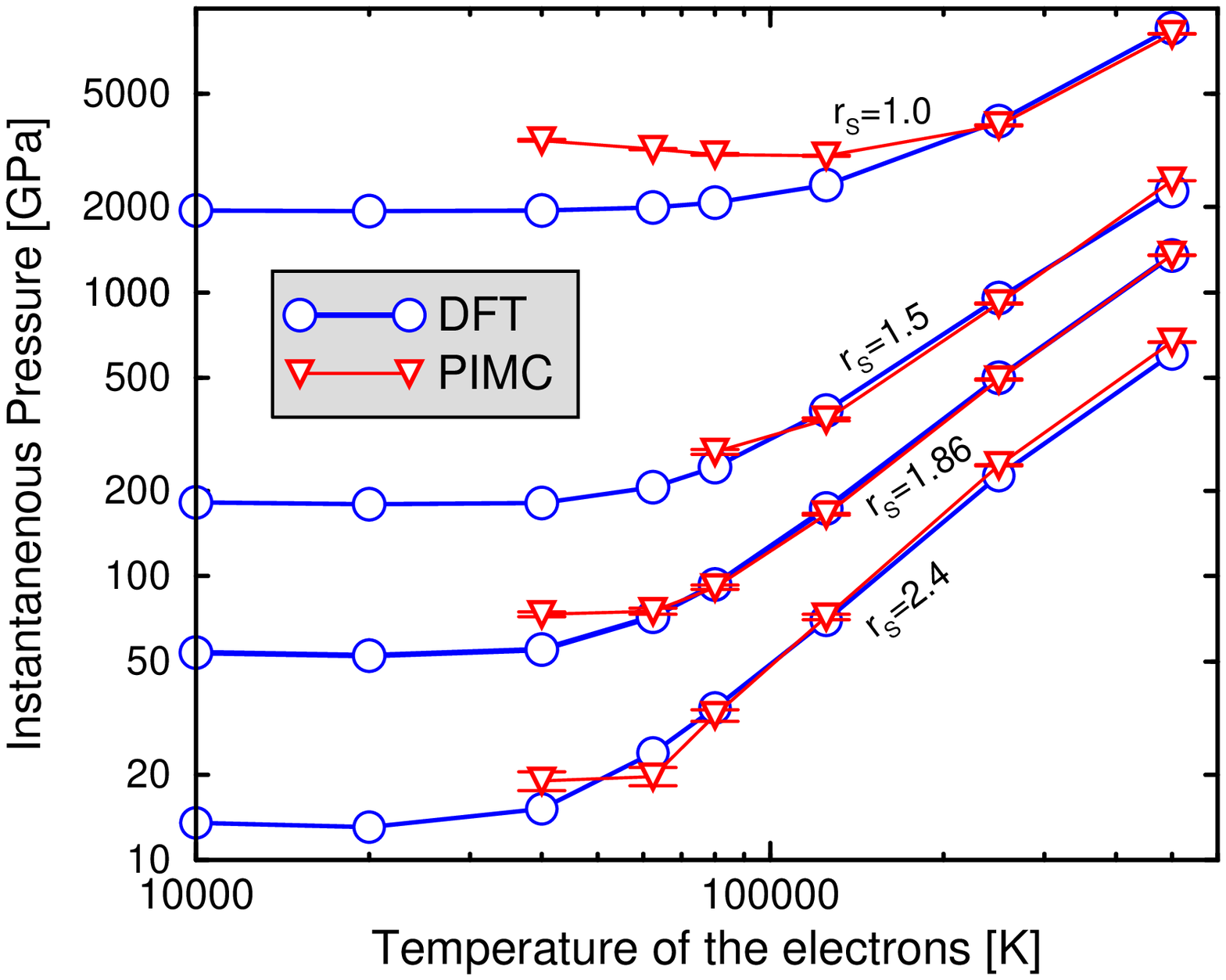}
\caption{ Comparison of the instantaneous pressure for a fixed 
        configuration of nuclei derived from PIMC and DFT with
        thermally populated states. The upper panel included results
        from PIMC with variational nodes. The lower panel compares
        PIMC results with free-particle nodes for different densities.
        (The ideal gas contributions from the nuclei are not
        included.) }
\label{Pinst}
\end{figure}

In order to explore the agreement between PIMC and DFT-MD in more
detail we resort to pressure calculations for a single configuration
of nuclei that we have obtain from DFT-MD with 57 atoms at
$80\,000\,$K. The nuclear coordinates are the $r_s=1.86$ snapshot are
given in Tab.~\ref{coord} in the appendix. For this fixed
configuration of nuclei, we now compare the instantaneous pressure as
a function of electronic temperature. The fact that the nuclei are now
classical rather being represented by paths in PIMC has a negligible
effect on the pressure for the temperatures under consideration. In
both methods, the instantaneous pressure is a well-defined quantity
derived from the virial theorem. In DFT, one uses the diagonal
elements of the stress tensor~\cite{StressNielsenMartin85}, while is
PIMC one derives the pressure directly from the kinetic,
$\left<\mathcal{K}\right>$, and potential, $\left<\mathcal{V}\right>$,
energy,
\beq
3 P V = 2 \left<\mathcal{K}\right> + \left<\mathcal{V}\right>\;,
\eeq
where $V$ is the volume of the simulation cell. DFT is primarily a
groundstate method that use an exchange functional that was derived at
$T$=0. It allows, however, to include electronic excitations using a
thermal population of excited singe-particle states. On the other
hand, PIMC is a finite temperature quantum simulation method that treats
the electrons as interacting particles. The only approximation comes
from the fermion nodes. 

Figure~\ref{Pinst} compares the instantaneous pressures from both
methods. At intermediate temperatures, there is a large interval where
both methods agree. DFT pressures appear to be fairly accurate. For
the level of accuracy needed for this study we could not detect any
insufficiency resulting from the groundstate exchange correlation
functional nor from inaccurate thermal excitations resulting from an
underestimated bandgap. However, the DFT eventually become
prohibitively expensive at higher temperature. Some of the points at
$r_s=2.4$ required up to 100 bands per atom, and that is one reason
why we only used a single configuration. The other comes from path
integrals. PIMC simulations with 123 atoms shown in Fig.~\ref{fig1}
represent about the limit one can study with currently available
computers. If one wants to repeat the calculations at lower temperature
where the paths are longer, or using the more expensive variational
nodes, one quickly exceeds existing limits in processing power.

Figure~\ref{Pinst} also reveals inaccuracies in the PIMC computation
that are caused by approximations in the trial density matrix. PIMC
with free-particle nodes predict pressures that are too high when the
electrons settle into the ground state ($T \le 40\,000\,$K for
r$_s$=1.86 as shown in Fig.~\ref{PPIMCDFT}). This effect has already
been reported for hydrogen~\cite{MC00}. In the variational density
matrix approach~\cite{MP00} one allows the trial density matrix to
adjust to the positions of the nuclei, which corrects most of the
pressure error as can be seen in upper panel of
Fig.~\ref{Pinst}. However, the variational approach was derived to
study finite temperature problems. It does not describe the electronic
ground state as well as DFT.

For the purpose of construction one EOS table for
helium, we use DFT-MD results with electronic excitation up to
$80\,000\,$K for $r_s\ge1.5$ and results up to $125\,000\,$K for $r_s
= 1.25$ and 1.0. For all higher $T$, we use PIMC simulations, which
become more and more efficient at higher $T$ because the length of the paths is
inversely proportional to temperature.

It should noted that the discussed validity range of different trial
density matrices depends very much on the material under
consideration. The temperature where we switch from PIMC to DFT-MD
reflects the degree of thermal electronic excitations as well as some
dependence of the approximations made in both methods. The density
dependence of the switching temperature would typically be estimated by
comparing the temperature to the Fermi energy of an ideal gas of
electrons. However, to incorporate band structure effects of dense
helium, we found it more appropriate to relate the switching
temperature to the DFT band width. Band width and Fermi energy are
identical in the systems of noninteracting particles. For the purpose
of this study, we found it appropriate to switch from PIMC to DFT-MD
for temperatures corresponding to less than one third of the helium
band width.

\begin{figure}[!]
\includegraphics[angle=0,width=\figurewidth]{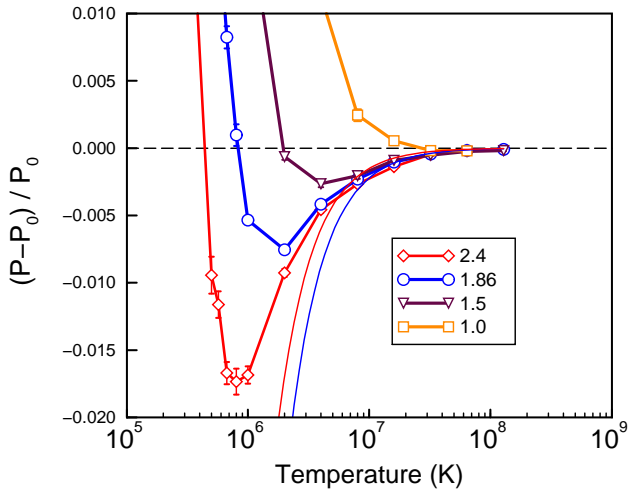}
\caption{ The relative excess pressure derived from PIMC (thick lines with symbols) 
        is compared with Debye-H\"uckel plasma model (thin lines) for
        different values of $r_s$ given in the legend. The ideal
        pressure, $P_0$, is derived Saha model of noninteracting
        helium species (see text).}
\label{debye}
\end{figure}

We performed PIMC simulations up to $1.28\times10^8\,$K covering a
large temperature interval of two orders of magnitude. The excess
internal energies and pressures in Figs.~\ref{PPIMCDFT},
\ref{EPIMCDFT}, and ~\ref{debye} are positive a lower temperatures
reflecting electronic excitations but then change sign due to
interactions of ions and free electrons. At very high temperature when
helium is fully ionized, the system can be described by the Debye
plasma model~\cite{DH23}. The Debye model is based on a self-consistent
solution of the Poisson equation for a system of screened charges. The
excess contribution to the free energy, internal energy, entropy per
particle, and pressure are given by,
\bea
\frac{F}{N_{\rm p}} = \frac{\Xi}{12}\;\;,\;\; 
\frac{E}{N_{\rm p}} = \frac{\Xi}{8}&,& 
\frac{S}{N_{\rm p}} = \frac{\Xi}{24}\;\;,\;\;
P = \frac{\Xi}{24 V}\;,
\\ 
\Xi= -k_B T V \frac{\kappa^3}{\pi}\;&,& 
\kappa^2 = \frac{4 \pi}{k_B T} \sum_i Z_i^2 \frac{N_i}{V},
\eea
where $\kappa=1/{r_d}$ is the inverse of Debye radius, $r_d$, in a
collection of $N_i$ particles of charge $Z_i$ in volume $V$ where
$N_{\rm p} = \sum_i N_i$. Figure~\ref{debye} demonstrate very good
agreement with the Debye model at high temperature. The Debye model
fails at lower temperatures where it predicts unphysically low
pressures. Under these conditions the screening approximation fails
because there are too few particles in the Debye sphere. The number of
particles in the Debye sphere is proportional to,
\beq 
(r_d/r_s)^3 \sim (T\,r_s)^{3/2}\;,
\eeq
which means that the Debye model becomes increasingly accurate for
high $T$ and large $r_s$. This is exactly what is observed in
Fig.~\ref{debye}. For higher densities, PIMC and Debye predictions
converge only at higher temperatures.

The size of the Debye sphere increases with temperature and will
eventually exceed the size of any simulation cell. This occurs when
the coupling the particles become very weak. With increasing
temperature, the Coulomb energy decreases while the kinetic energy
increase linearly with $T$. To determine the precise amount of the
Coulomb energy becomes more and more difficult due to finite size
effects. This is the reason why the PIMC simulations do not agree
perfectly with the Debye model for the highest temperature shown in
the lower panel of Fig.~\ref{EPIMCDFT}. In conclusion, we use the
Debye EOS for the highest data in our EOS table. The nonideal
pressures reported in Fig.~\ref{debye} appear to be less sensitive to
finite size errors than the internal energy because their volume
dependence is relatively weak.

\begin{figure}[!]
\includegraphics[angle=0,width=\figurewidth]{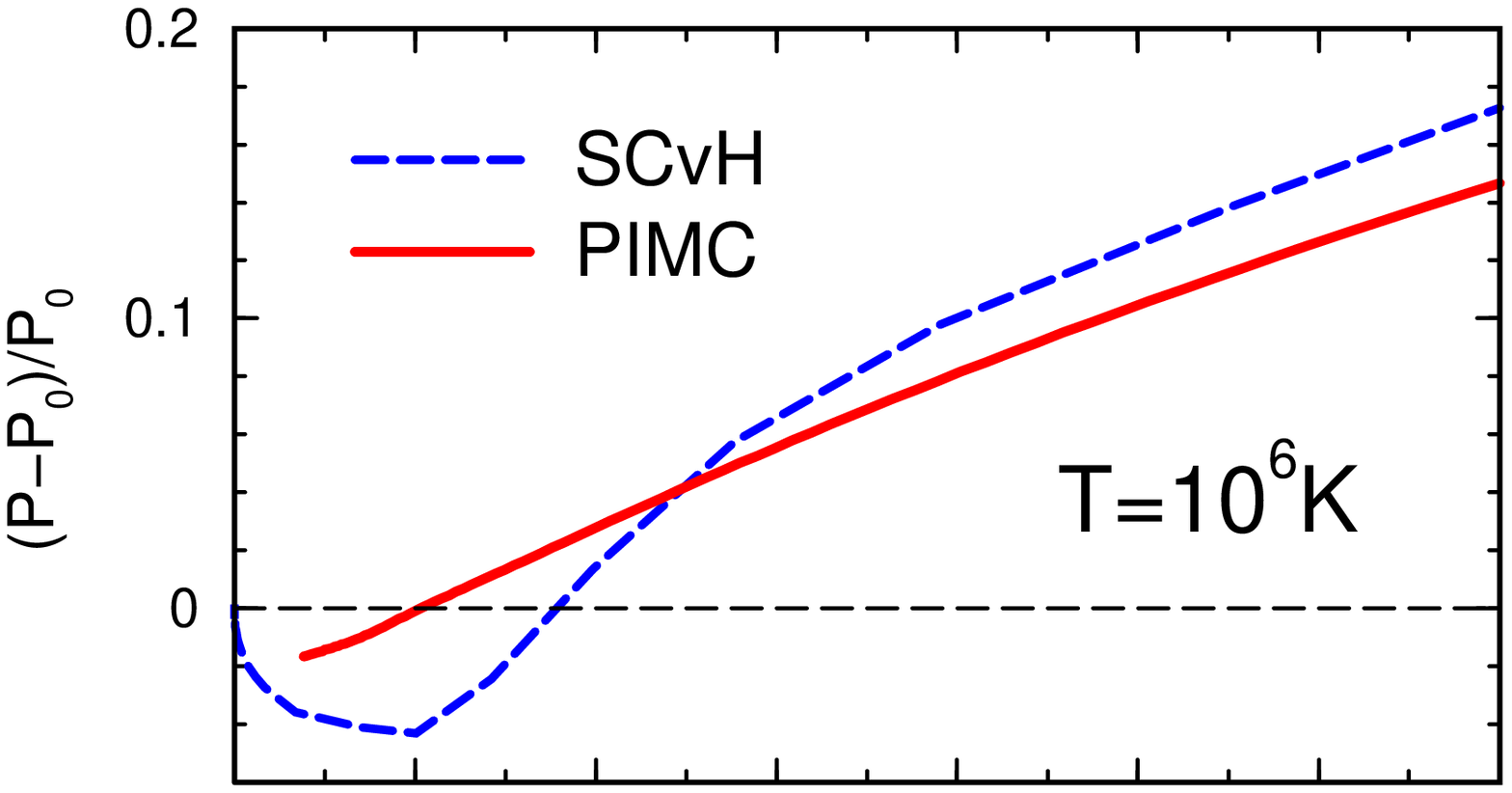}
\includegraphics[angle=0,width=\figurewidth]{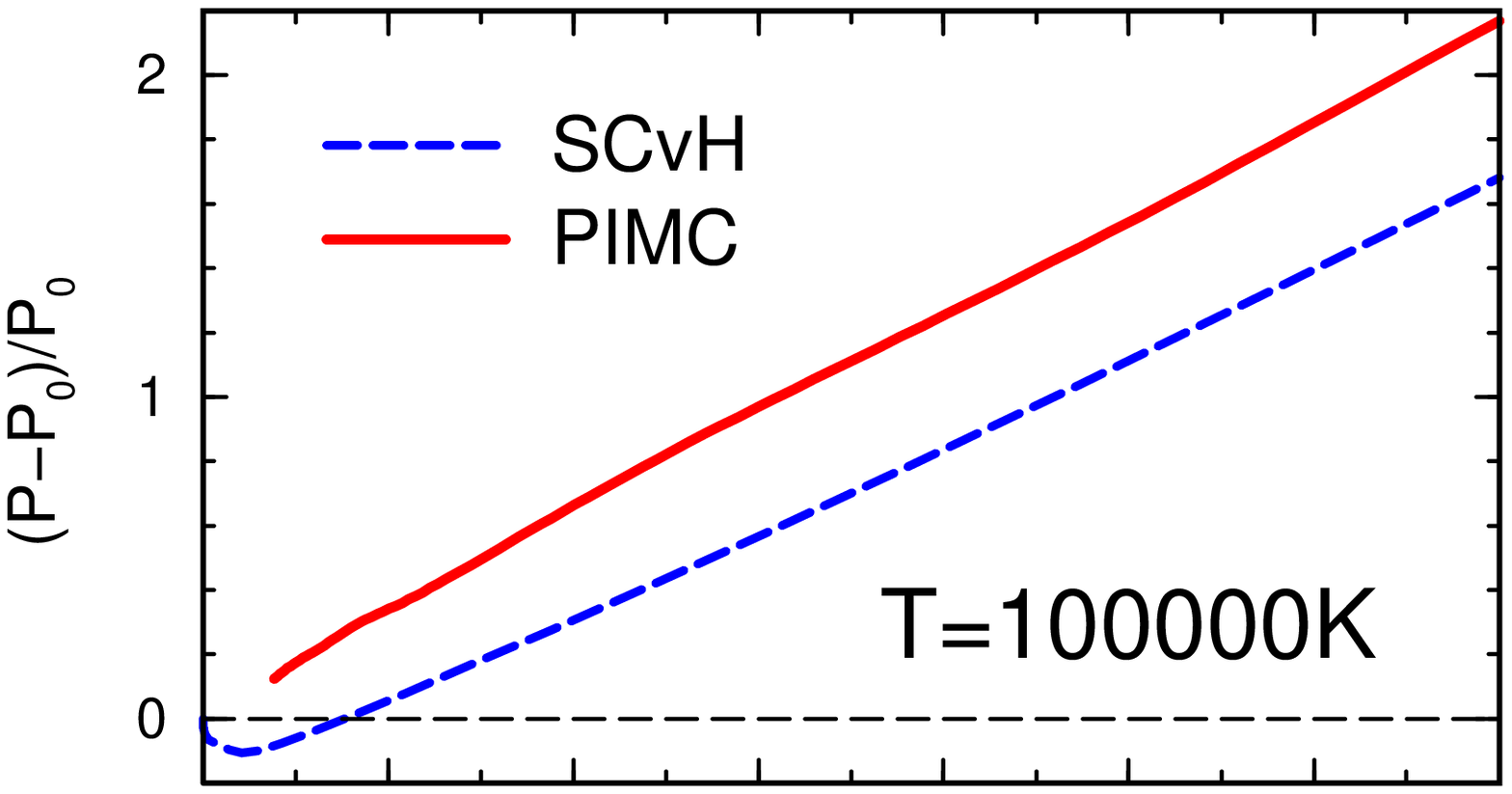}
\includegraphics[angle=0,width=\figurewidth]{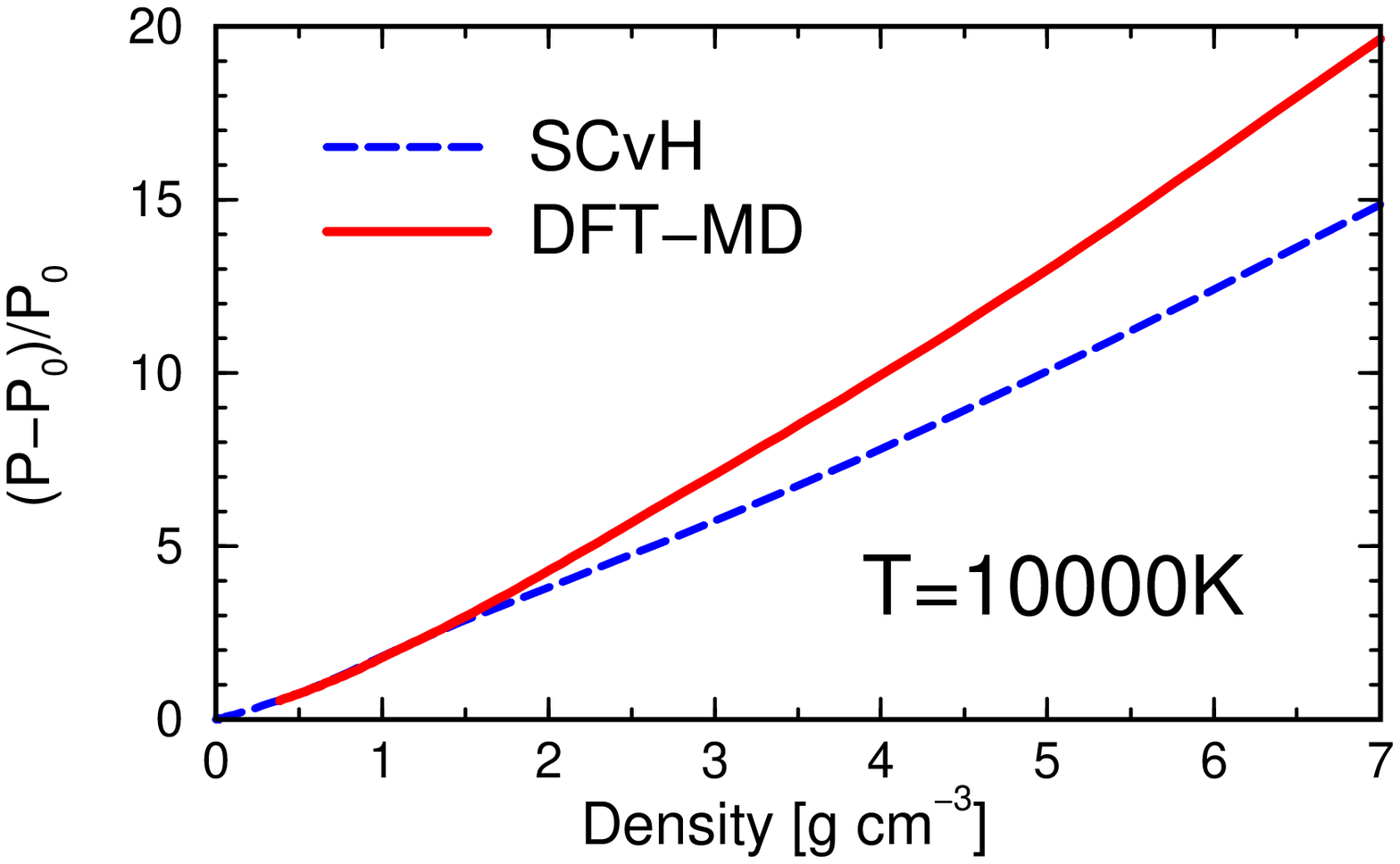}
\caption{ Comparison of the relative excess pressure 
	reported by first-principles simulations with the SCvH EOS
	model.  The three temperatures shown here are relevant for
	stellar interiors, the comparison with shock wave experiments,
	and the interiors of giant planets.}
\label{PcompSC}
\end{figure}

Now we compare our first-principle EOS with chemical free energy
models that were developed before first-principles simulation data
became available. Winisdoerffer and Chabrier~\cite{WC05} constructed a
semianalytical model to study stellar interiors that covers a wide
density range including metallization. Their EOS is only available in
explicit form in a small temperature interval and a comparison with
DFT-MD simulation has already been reported~\cite{Kietzmann07}. That
is why we focus on the free energy model derived by Saumon, Chabrier,
and van Horn (SCvH)~\cite{SC95}. Together with their hydrogen
model~\cite{SC92}, their EOS has been used numerous times to model
giant planet interiors.

Figure~\ref{PcompSC} compares the excess pressure from both EOSs for
three different temperatures. At a very temperature of 10$^6$K that
are important for stellar interiors, we found fairly good
agreement. The deviations between the SCvH model and PIMC simulations
are only about 4\%. 

At a intermediate temperature of $100\,000\,$K, that approximately
represent the regime of shock wave experiments, the agreement is less
favorable. One finds that the SCvH EOS reports pressures that are
about 30\% lower than those predicted by PIMC. This is partly
due to the fact that the SCvH model follows the Debye model down to
too low temperatures (Fig.~\ref{EPIMCDFT}) and that an interpolation
scheme between a low and a high-temperature model was used by SCvH
that was not thermodynamically consistent (Fig.~\ref{hug1}).

The last panel in Fig.~\ref{PcompSC} is focused on the interiors of
giant planets with temperature of the order to 10$\,000\,$K. At low
density both EOSs agree well, but above 1.5$\,$g$\,$cm$^{-3}$
deviation begin to increase steadily. At conditions comparable to
Jupiter's interior, we find that the SCvH underestimates the pressure
by 30\%. In a hydrogen-helium mixture of solar composition, this
translates into an error in the pressure of about 4\%. This is the
reason why even the helium EOS is important to estimate the size of
Jupiter's core, which is expected to be only a small fraction of the
Jupiter's total mass.

\section{Pair correlation functions}

In this section, we study the structure of the fluid by analyzing
correlations between different types of particles. Given the large
amount of simulation results, we focus our attention primarily on the
temperature dependence and only report results for one density of
$r_s=1.86$. The density dependence of the pair correlation functions,
$g(r)$, has been analyzed in Ref.~\cite{MC01}.

\begin{figure}[!]
\includegraphics[angle=0,width=\figurewidth]{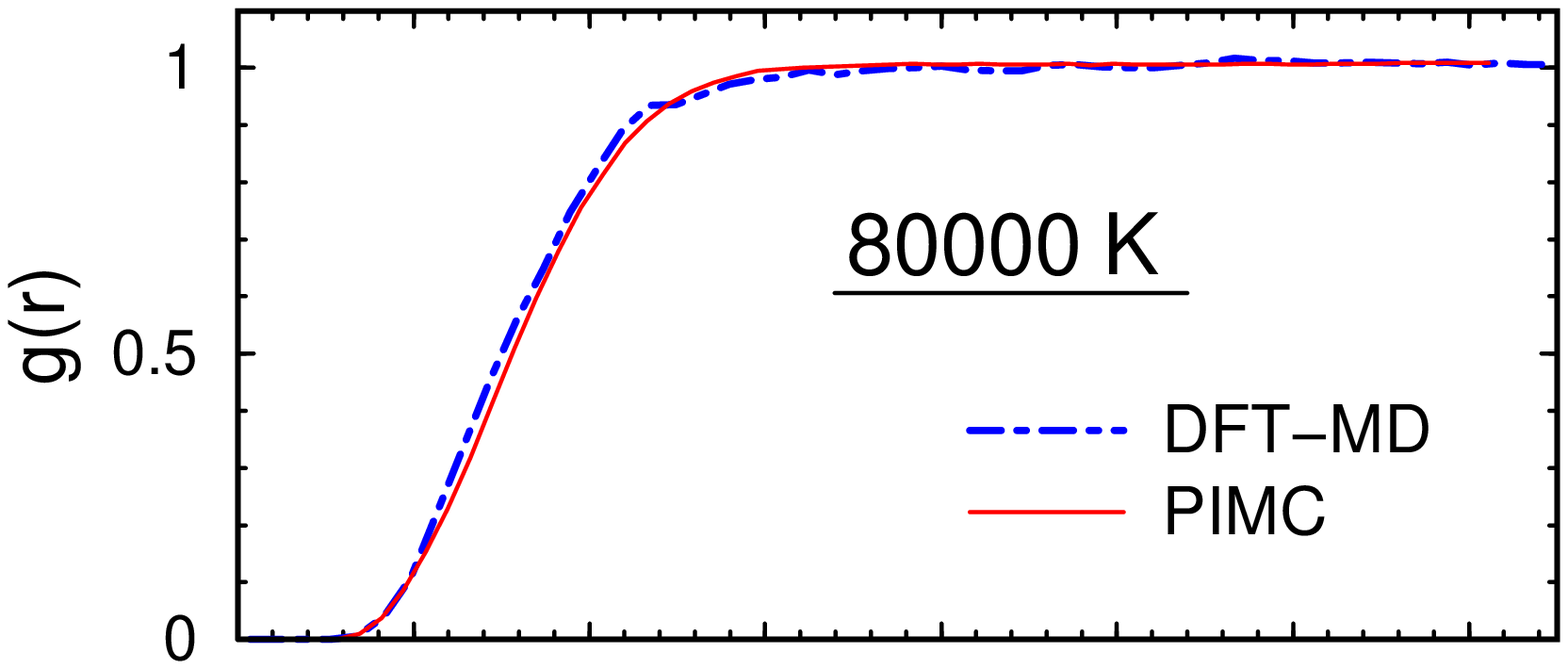}
\includegraphics[angle=0,width=\figurewidth]{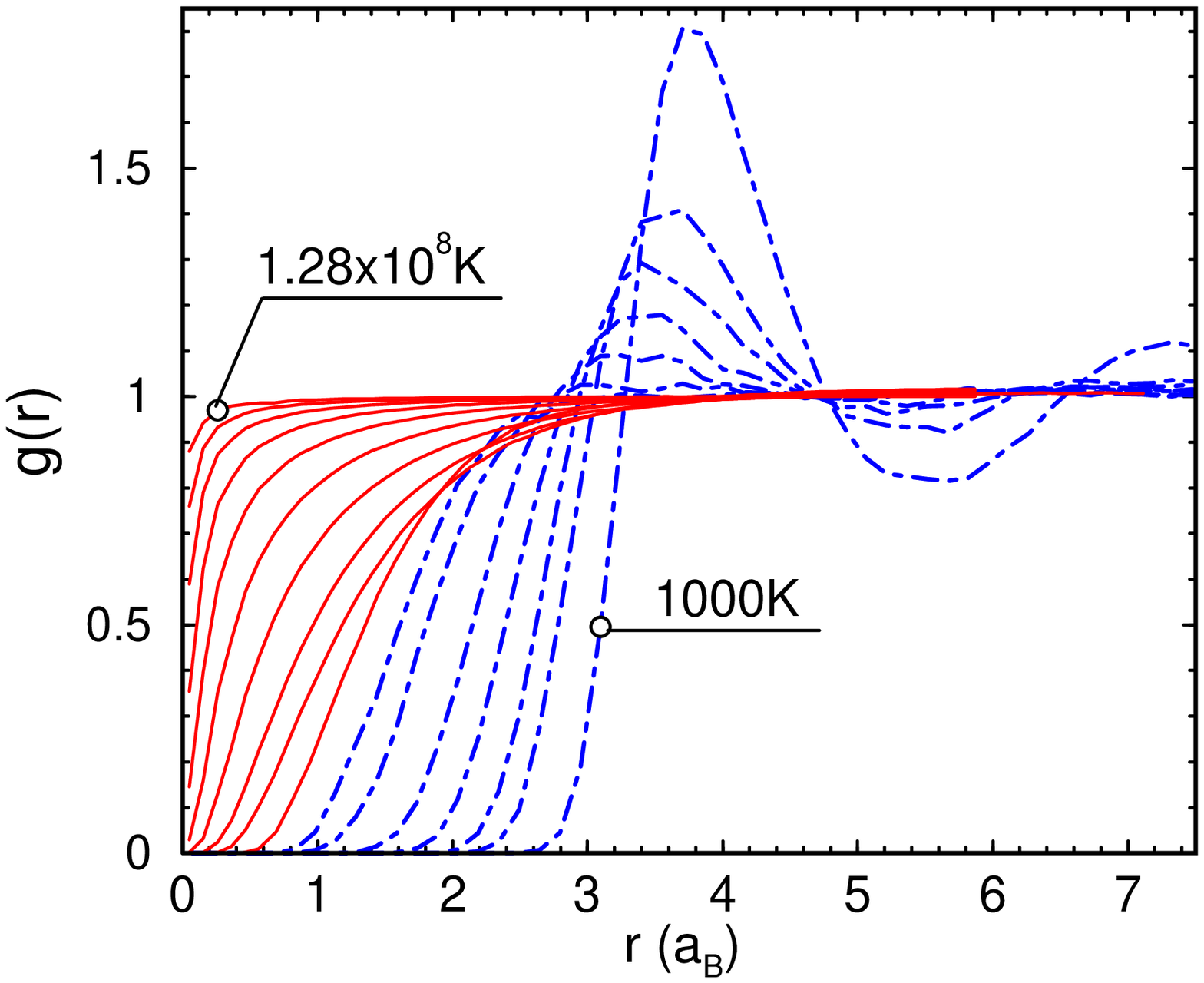}
\caption{The nuclear pair correlation functions for $r_s=1.86$. The
       lower panel shows the following temperatures: \{ 128, 64, 32, 16,
       8, 4, 2, 1, 0.5, and 0.125 \} $\times 10^6\,$K from PIMC as well
       as \{ 40, 20, 10, 5, 3, and 1 \} $\times 10^3\,$K 
       from DFT-MD.}
\label{fig_graa}
\end{figure}

Figure~\ref{fig_graa} shows how the nuclear pair correlation functions
changes over a temperature interval that spans seven orders of
magnitude. At low temperature, the $g(r)$ shows the oscillatory
behavior that is typical for a hard-sphere fluid. The atomic
interactions are governed by two tightly bound electrons that lead to
a strong repulsion at close range due to Pauli exclusion. As long as
the density is not too high, this behavior is well-described by the
Aziz pair potential~\cite{Mi06}.

As temperature increases, two effects change the pair correlation
function. The increase in kinetic energy leads to stronger collisions
and atoms approach each other more. This is regard, helium is not
exactly a hard-sphere fluid because the Aziz pair potential is not
perfectly hard. The increase in temperature also damps of the
oscillation in the $g(r)$.

At 80$\,$000 K, one finds perfect agreement between PIMC and DFT-MD
(upper panel in Fig.~\ref{fig_graa}). As the temperature is increased
further, changes in nuclear $g(r)$ function are dominated by
electronic excitations and the ionization of atoms. One finds that the
strong repulsion at low temperature disappears gradually. As one
approaches the Debye-H\"uckel limit, the fluid behaves like correlated
system of screened Coulomb charges.

\begin{figure}[!]
\includegraphics[angle=0,width=\figurewidth]{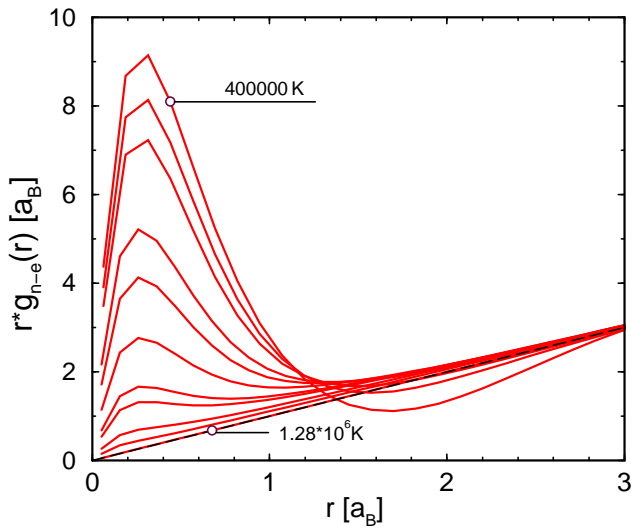}
\caption{The electron-nucleus pair correlation functions, $g(r)$, from 
       PIMC for $r_s=1.86$. Starting with the highest peak, the
       following temperatures are plotted: \{ 0.04, 0.08, 0.125, 0.25,
       0.333, 0.5, 0.8, 1, 2, 4, and 128 \} $\times 10^6\,$K. We plot
       $r*g(r)$ on the ordinate so that the peak at small $r$
       illustrates the fraction of electrons in bound states. The
       decrease in peak height with increasing temperature
       demonstrates thermal excitation of electrons, which eventually
       leads the ionization of atoms. }
\label{fig_grae}
\end{figure}

The peak in the electron-nucleus pair correlation functions in
Fig.~\ref{fig_grae} illustrates that the electrons are bound to the
nuclei. At 40$\,$000 K and below, the peak height is maximal. At
higher temperature, electrons get excited and eventually atoms get
ionized. The peak height is consequently reduced until, at very high
temperature, the motion of electrons and nuclei appears to be
uncorrelated.

\begin{figure}[!]
\includegraphics[angle=0,width=\figurewidth]{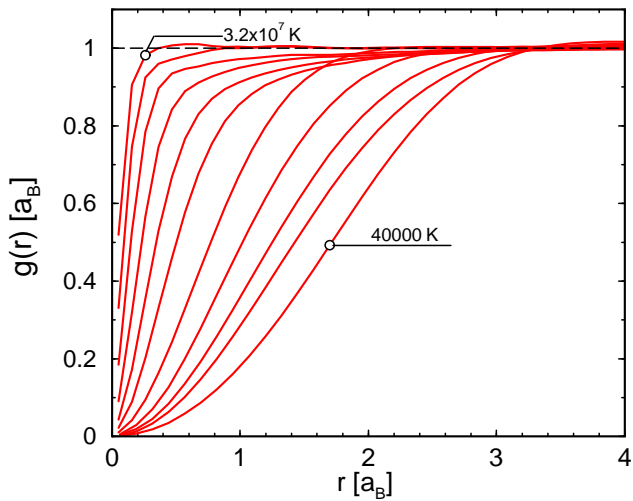}
\caption{The electron-electron pair correlation functions for electrons 
       with parallel spins calculated with PIMC for
       $r_s=1.86$. Starting from the left, the following temperatures
       are plotted: \{ 32, 16, 8, 4, 2, 1, 0.25, 0.125, 0.08, 0.0625,
       0.04 \} $\times 10^6\,$K. With increasing temperature, correlation
       effects are reduced and the exchange-correlation hole
       disappears.}
\label{fig_gree_l}
\end{figure}

The correlation of electrons with parallel spins is determined by
Pauli exclusion and Coulomb repulsion but is also influenced by the
motion of the nuclei at low temperature. Combination of all these
effects causes the motion of same-spin electrons to be negatively
correlated at small distances. This is typically referred to as the
exchange-correlation hole. At high temperatures, kinetic effect
reduces the size of this hole but $g(r)$ always goes to zero for small
$r$ due to Pauli exclusion.

\begin{figure}[!]
\includegraphics[angle=0,width=\figurewidth]{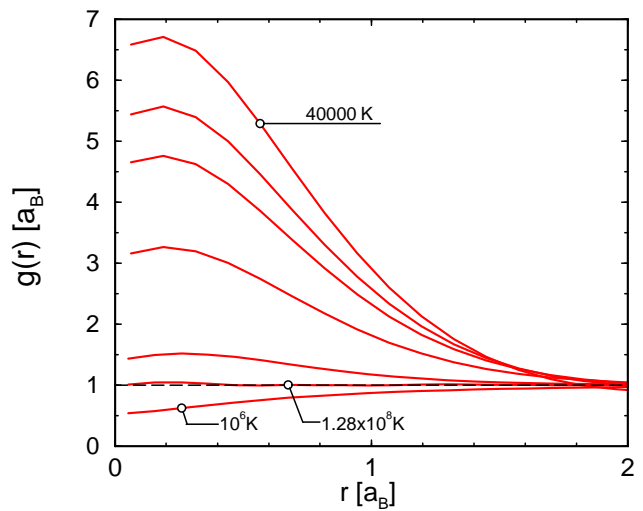}
\caption{The electron-electron pair correlation functions for electrons 
       with opposite spins calculated with PIMC for
       $r_s=1.86$. Starting from the top, the following temperatures
       are plotted: \{ 0.04, 0.0625, 0.08, 0.125, 128, and 1 \} $\times
       10^6\,$K. The smallest values are observed for $10^6\,$K. }
\label{fig_gree_u}
\end{figure}

Despite the Coulomb repulsion, the electrons with opposite spins are
positively correlated at low temperature, because two electrons with
opposite spin are bound in a helium atom. With increasing temperature,
the peak in Fig.~\ref{fig_gree_u} reduces in height because more and
more electrons get ionized. At 10$^6$, one finds the lowest values for
$g(r\to 0)$ because the electrons are anti-correlated due to the
Coulomb repulsion. If the temperature is increased further kinetic
effects dominate over the Coulomb repulsion and $g(r\to 0)$ again
increases and will eventually approaches 1 at high temperature.

\section{Entropy calculations}

\begin{figure}[!]
\includegraphics[angle=0,width=\figurewidth]{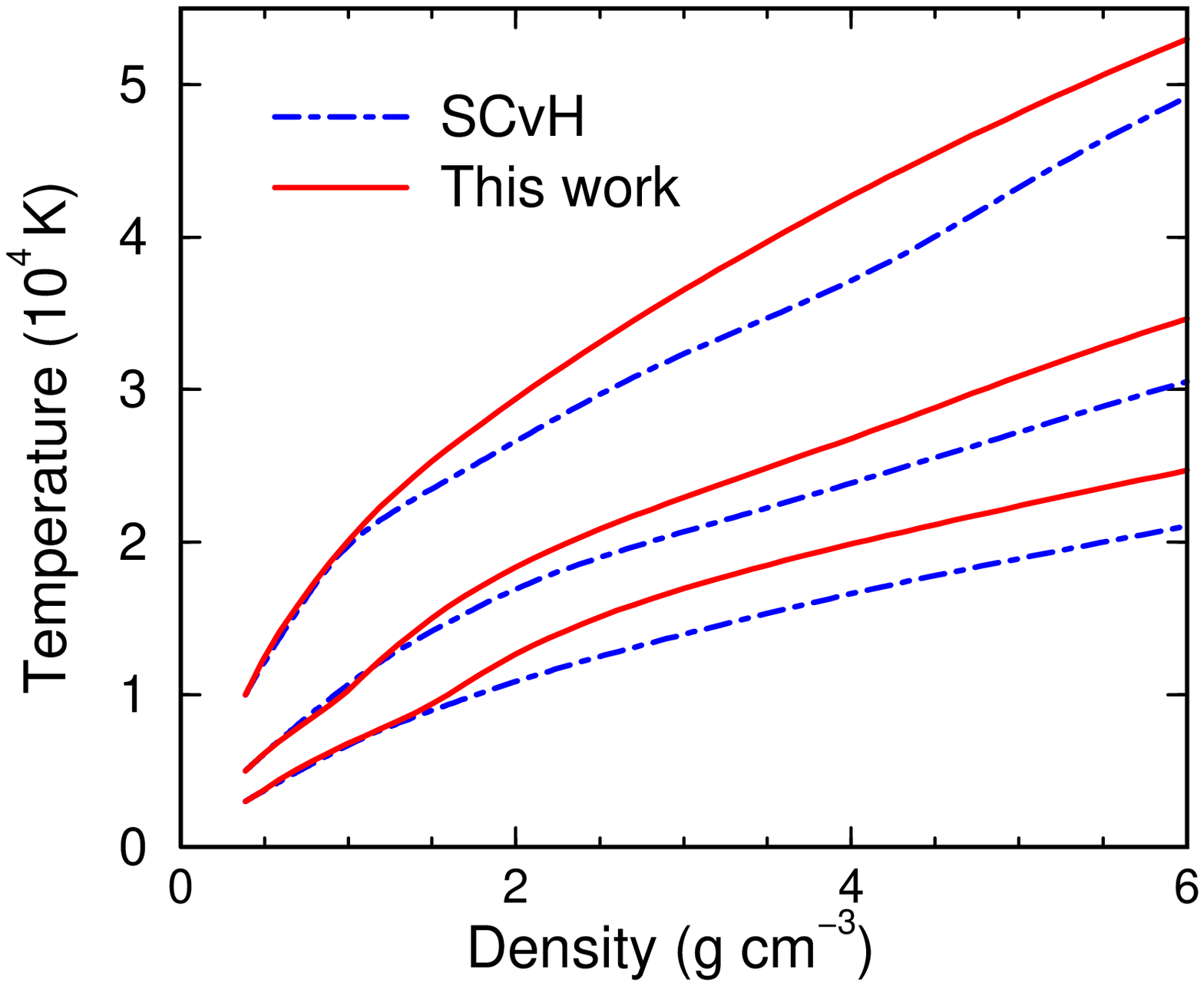}
\includegraphics[angle=0,width=\figurewidth]{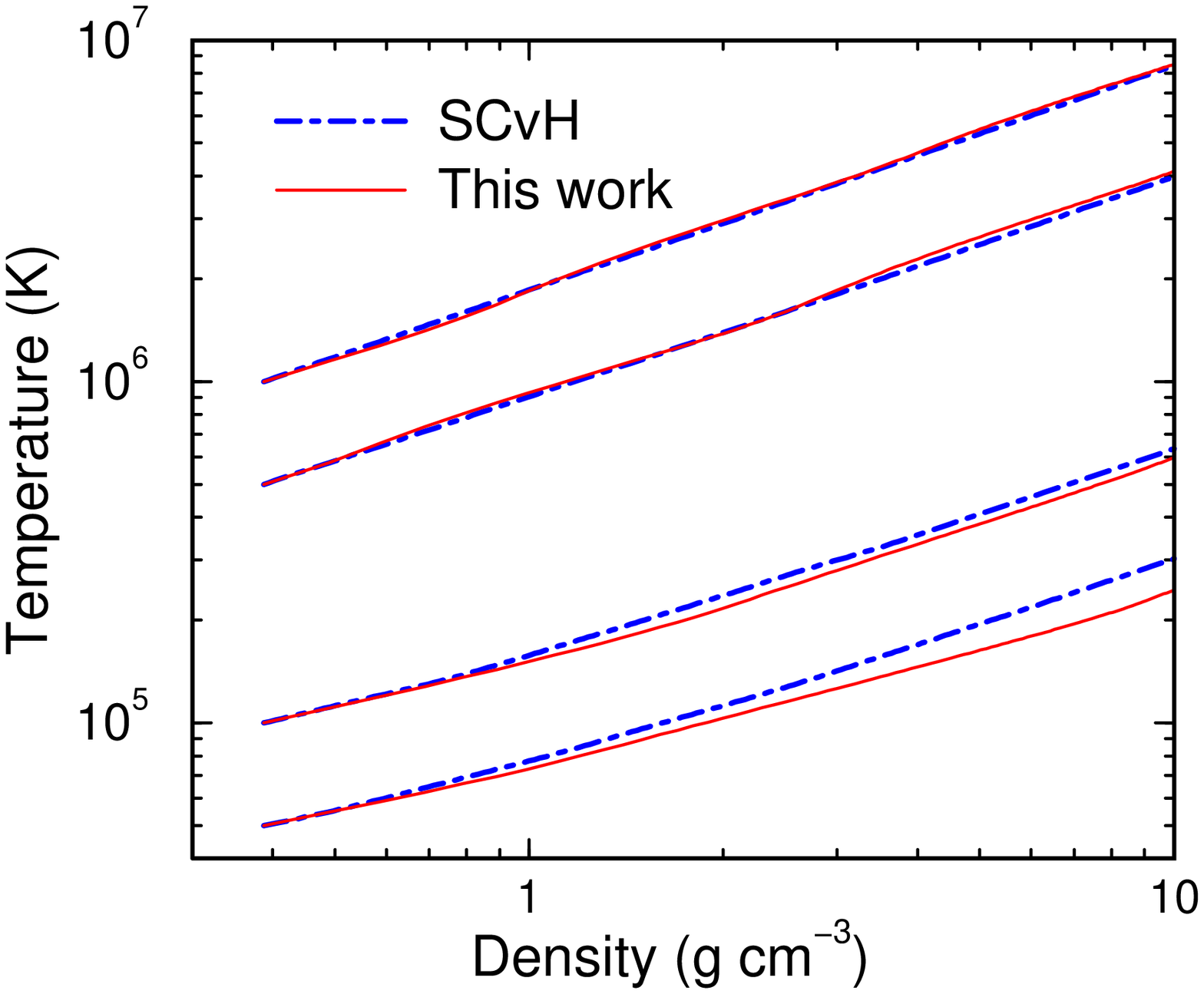}
\caption{ Comparison in temperature-density space of adiabats
         from first-principles simulations (this work) and the SCvH
         EOS model. }
\label{ad1}
\end{figure}

\begin{figure}[!]
\includegraphics[angle=0,width=\figurewidth]{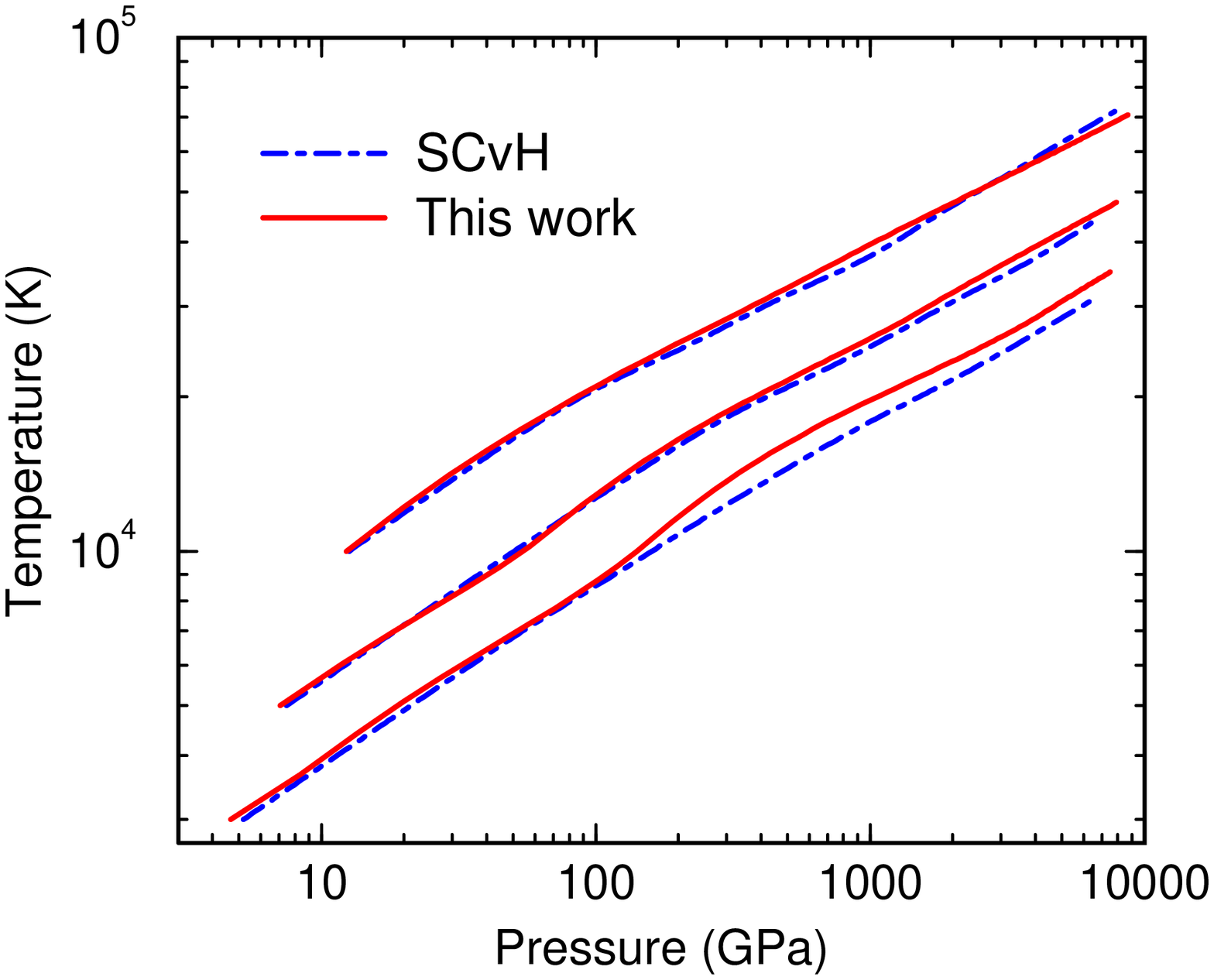}
\includegraphics[angle=0,width=\figurewidth]{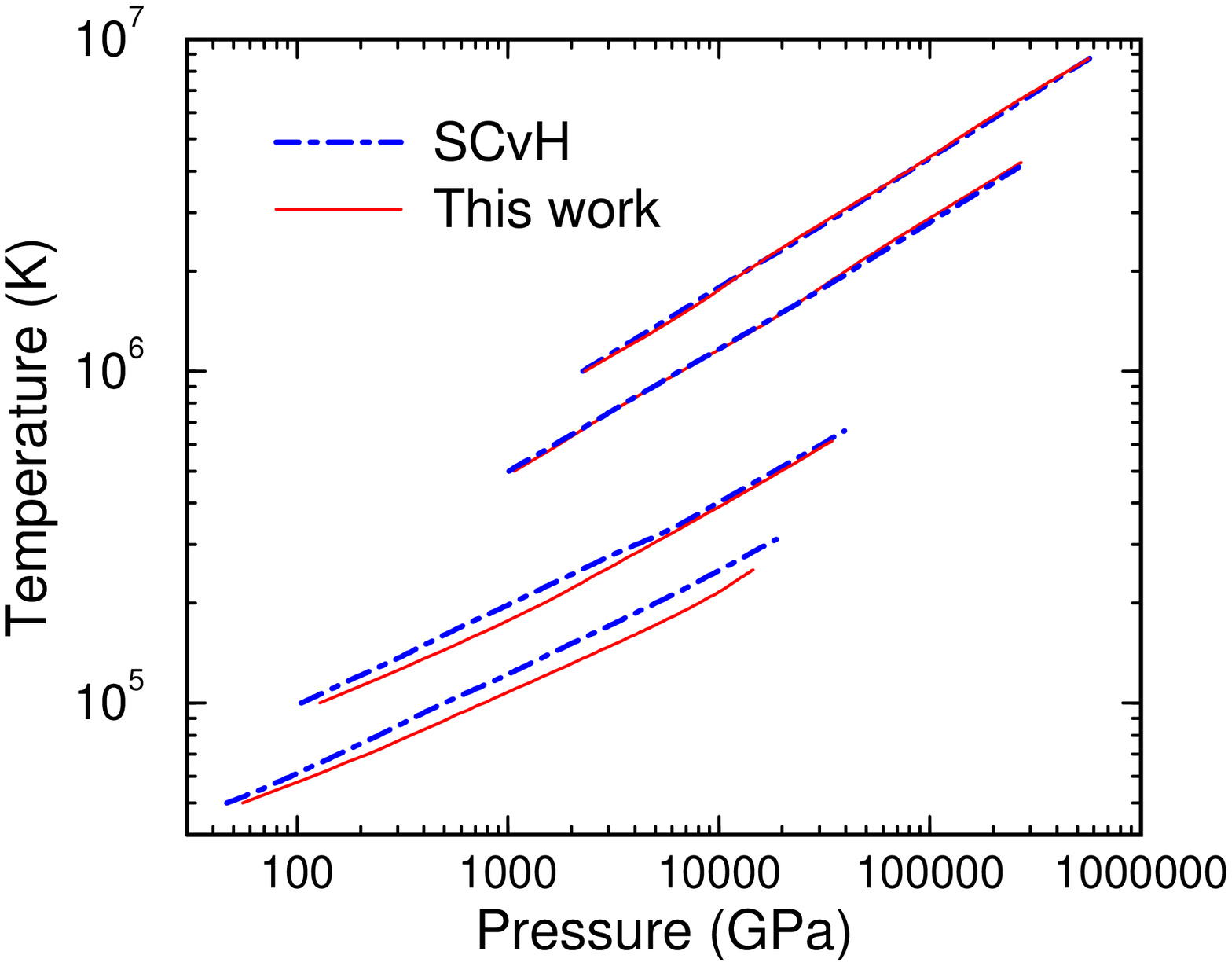}
\caption{ Comparison in temperature-pressure space of the adiabats shown in Fig.~\ref{ad1}. }
\label{ad2}
\end{figure}

Convection in the interior of planets determines that the
temperature-pressure profile is adiabatic. In consequence, the
planetary interiors is fully determined by the conditions on the
surface and the EOS. This makes the calculation of adiabats
important. However, neither Monte Carlo nor molecular dynamics methods
can directly compute entropies because both techniques save orders of
magnitude in computer time by generating a only representative sample
of configurations. Without this gain many-body simulations would be
impossible and in consequence, entropies that are measure of the total
available phase space are not accessible directly. 

Typically one derives the entropy by thermodynamic integration from a
know reference state. However, for the planetary interiors, the
absolute value of the entropy is not important as long as one is able
to construct $(T,P)$ curve of constant entropy. This can be done using
the pressure and the internal energy from first-principles simulations
at different $(T,V)$ conditions. Using Maxwell's relations, one finds,
\beq
\left. \frac{\partial T}{\partial V}\right|_S = - \frac
{ \left. \frac{\partial S}{\partial V}\right|_T }
{ \left. \frac{\partial S}{\partial T}\right|_V } 
= - T \frac
{ \left. \frac{\partial P}{\partial T}\right|_V }
{ \left. \frac{\partial E}{\partial T}\right|_V } \;.
\label{entropy}
\eeq
By solving this ordinary differential equation, (V,T)-adiabats can be
constructed as along as a sufficiently dense mesh of high-quality EOS
points are available to make the required interpolation and
differentiation of $E$ and $P$ with respect to temperature
satisfactorily accurate.

One drawback of formula~(\ref{entropy}) is that it is not necessarily
thermodynamically consistent if pressures and internal energies are
interpolated separately. This is the primary reason why we developed the
following method to fit the free energy instead. Pressure and internal
energy are related to the free energy, $F(V,T)$, by
\beq
P =   -\left. \frac{\partial F}{\partial V}\right|_T \;\;\;\;\;\;{\rm and}\;\;\;\;\;\;
E = F -T \left. \frac{\partial F}{\partial T}\right|_V\;.
\eeq

Different EOS fits for fluids have been proposed in the
literature~\cite{Le00,StringfellowDeWittSlattery90}. Thermodynamic
consistency was not a priority in each case. Both papers relied on
specific functional forms that were carefully adjusted to the material
under consideration. Although such a fit of specific form could
probably have also been constructed for the presented helium EOS data,
we wanted to have an approach that is not just applicable to one
material. Therefore, we decided to represent the free energy as a
bi-cubic spline function with temperature and density as
parameters. This spline function can accurately represent our helium
EOS data and can easily be adapted to fit other
materials. Cubic splines are twice continuously differentiable, which
means the derived pressures and energies are once continuously
differentiable with respect to $V$ and $T$. This is sufficient for
this study. If additional thermodynamic functions that require higher
order derivatives of the free energy, such as sound speeds, need to be
fit also then higher order splines can accommodate that.

We start the free energy interpolation by constructing as series of
one-dimensional splines functions $F_n(T)$ for different
densities. The choice of knots $T_i$ is arbitrary. Their location
should be correlated with the complexity of the EOS as well as the
distribution of EOS data points. In our helium example, we used a
logarithmic grid in temperature with about half as many knots as data
points. The set of free energy values on the knots, $F(T_i)$,
represent the majority of the set of fit parameters. In addition, one
may also include the first derivatives of the splines $\frac{\partial
F_n}{\partial T}|_V$ at the lowest and highest temperature, which
represent the entropy. Alternatively, one could derive those
derivatives by other means and then keep them fixed during the fitting
procedure.

To compute the free energy at a specific density, $n^*$, and
temperature, $T^*$, we first evaluate all splines $F_n(T^*)$ and then
construct a secondary spline at constant temperature as function of
density, $F_{T^*}(n)$. Its first derivate is related to the
pressure. Again, the derivative at the interval boundaries can either
be fixed or adjusted during the fitting procedure. We adjust them by
introducing an additional spline $\frac{\partial F}{\partial n}|_T(T)$
at the lowest and highest densities, which then get adjusted in the
fitting procedure.

We begin the fitting procedure with an initial guess for the free
energy function derived from Eq.~(\ref{entropy}). Then we employ
conjugate gradient methods~\cite{numerical_recipes} to optimize the
whole set of fitting parameters. Minimizing the sum of the squared
relative deviations in pressure and internal energy has been found to
work best. (Just for the derivation of the relative deviation in
energy, the zero of energy has been shifted to value of the isolated
helium atom.)

All fits tend to introduce wiggles if too many free parameters are
included. We control wiggles by adjusting the number of knots in
density and temperature but we also introduced penalty in the form,
\beq
\xi = \int d\rho \left( \frac {\partial^3 F}{\partial n^3} \right)^2\;\;,
\eeq
to favor fits with smaller $|\partial^2 P / \partial n^2|$. Finally we
changed the density argument in the spline interpolation from
$F_T(\rho)$ to $F_T(\log(\rho))$. This improves the fit in the high
temperature limit where the free energy is dominated by the ideal gas
term that has logarithmic dependence on density.

The presented free energy fit is thermodynamically consistent by
construction. It allowed us to accurately represent the entire data
set of $P$ and $E$ values. Without additional information, the free
energy can be determined up to a term $T\Delta S$, which is sufficient
to compute adiabats. To determine the absolute value of the entropy,
one needs an anchor point, for which the entropy was derived by
different means.

Figure~\ref{ad1} compares different adiabats derived from our
first-principles EOS with predictions from the SCvH EOS
model. Beginning from a joint starting point of $r_s=2.4$ and a
selection of seven different temperatures of 3000, 5000, 10$\,$000,
50$\,$000, 100$\,$000, 500$\,$000, and 10$^6\:$K, we constructed the
adiabats for both models for the density interval under
consideration. The upper panel of Fig.~\ref{ad1} demonstrates good
agreement between both methods a low densities up to about 1 g
cm$^{-3}$. For higher densities, one finds deviations of up to 20\% in
the predicted temperatures on the adiabat. A higher temperature, the
agreement get substantially better, which is illustrated in the lower
panel of Fig.~\ref{ad1}. The observed deviations are similar to
pressure differences shown in Fig.~\ref{PcompSC}.

For applications in the field of planetary science, we also show the
adiabats in ($T$,$P$) space in Fig.~\ref{ad2}. The deviations are
comparable in magnitude but appear smaller on a logarithmic scale.

\section{Shock wave experiments}

\begin{figure}[!]
\includegraphics[angle=0,width=\figurewidth]{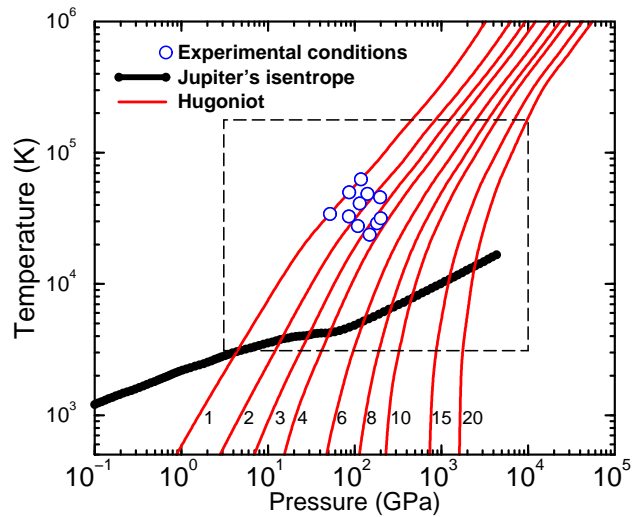}
\caption{Comparison of Jupiter's isentrope with helium shock 
       Hugoniot curves for different precompression ratios. The labels
       specify the precompression ratio relative to the density at
       ambient pressures ($\rho_0=0.1235\,$g$\,$cm$^{-3}$,
       Ref.~\cite{nellis84}). The symbols approximately represent
       recent experiments~\cite{eggert08}. The inside of the dashed
       box indicates conditions, for which the SCvH EOS~\cite{SC95} was
       interpolated and is not thermodynamically consistent.}
\label{hug1}
\end{figure}

\begin{figure}[!]
\includegraphics[angle=0,width=\figurewidth]{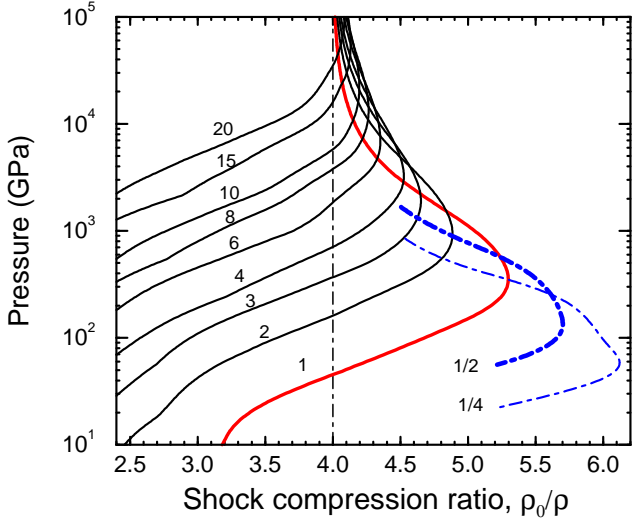}
\caption{ Hugoniot curves for different precompression ratios 
	from Fig.~\ref{hug1} plotted as function of shock compression.}
\label{hug2}
\end{figure}

\begin{figure}[!]
\includegraphics[angle=0,width=\figurewidth]{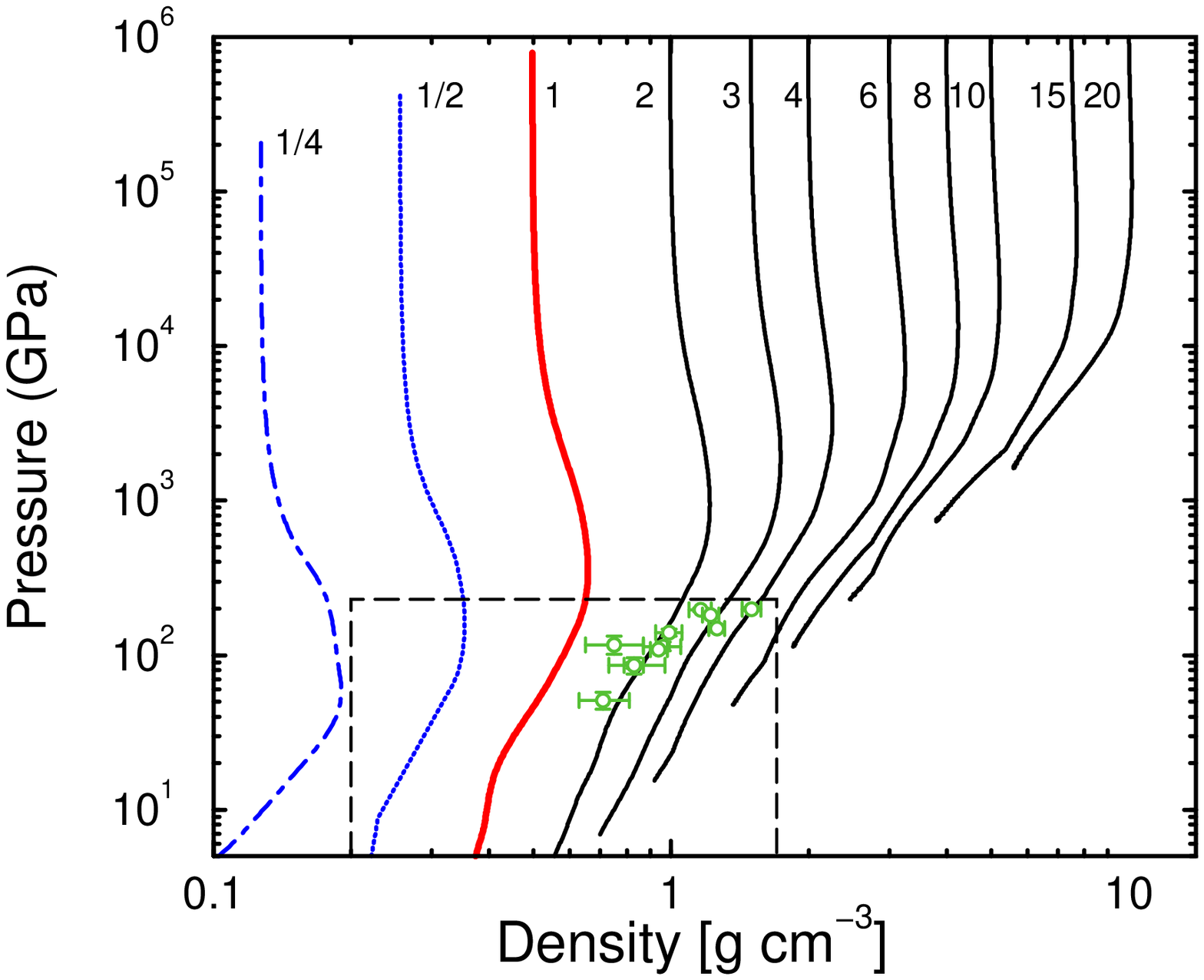}
\includegraphics[angle=0,width=\figurewidth]{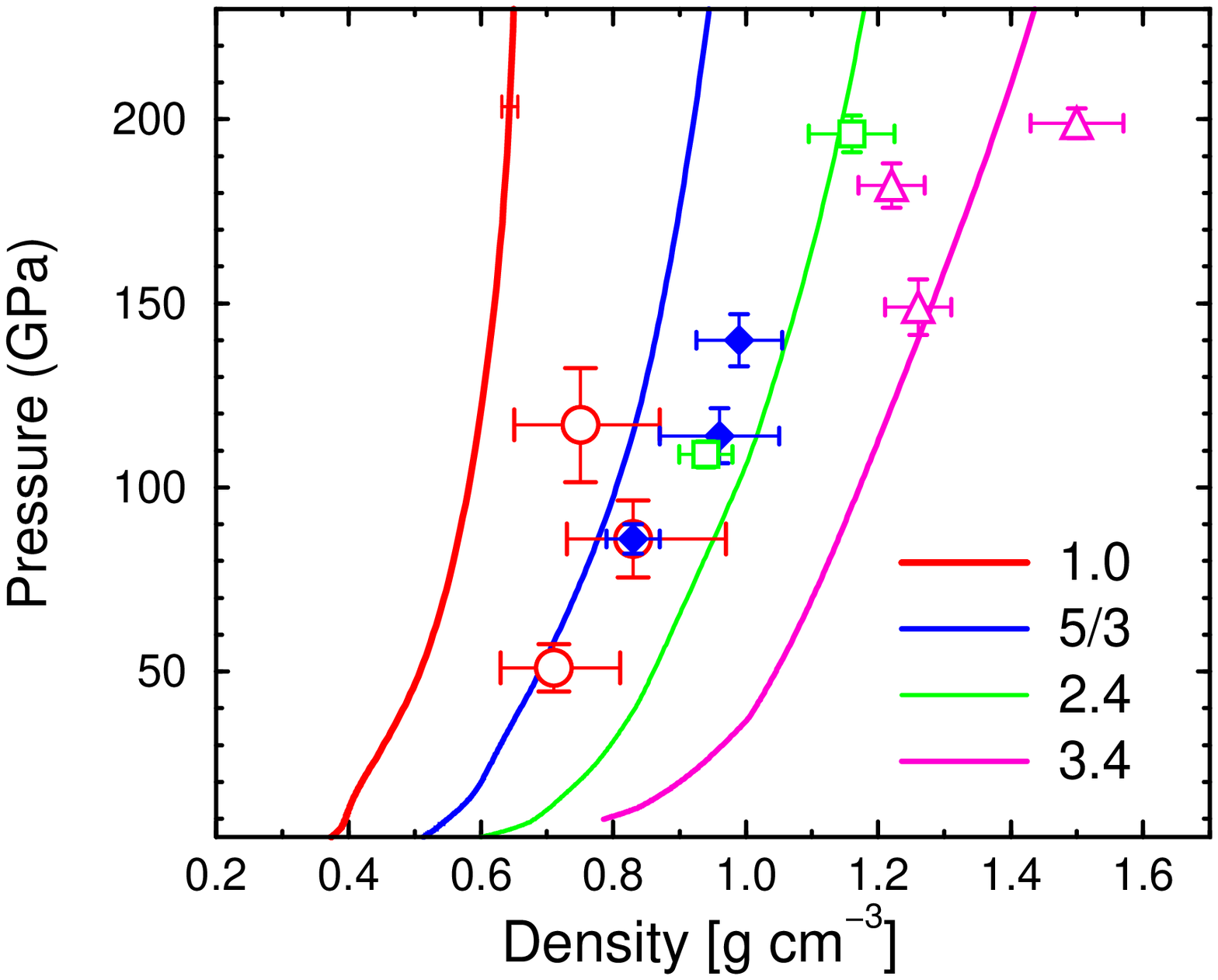}
\caption{ Pressure-density plot of shock Hugoniot curves for different 
        precompression ratios. The upper panel shows the shock
        Hugoniot curves from Fig.~\ref{hug2}, recent experimental
        results (symbols), and the range of the lower plot. Below we
        plot the Hugoniot curves for precompression ratios (see
        labels) that approximately match the experimental conditions
        (symbols, see Ref.\cite{eggert08}). The error bar on the
        upper left solid line represents the uncertainty in the
        calculations.}
\label{hug3}
\end{figure}

Dynamic shock compression experiments are the preferred laboratory
experiments to probe the properties materials at high pressure and
temperature. Using bigger and more powerful lasers, great progress has
been made in reaching higher and higher pressures. Under shock
compression, the initial state of material characterized by internal
energy, pressure, and volume ($E_0,P_0,V_0$) changes to the final
state described by $(E,P,V)$. The conservation of mass, momentum, and
energy yields the Hugoniot condition~\cite{Ze66},
\beq
H = (E-E_0) + \frac{1}{2} (P+P_0)(V-V_0) = 0.
\label{hug}
\eeq
Different shock velocities lead to a collection of final states that
are described by a Hugoniot curve. Using Eq.~\ref{hug}, this curve can
easily calculated for a given EOS where one most often may assume $P_0
\ll P$. $V_0=32.4$ {\rm cm}$^3$/mol ($\rho_0=0.1235\,$g$\,$cm$^{-3}$) is 
taken from experiment~\cite{nellis84}. For $E_0$, one takes the energy
of an isolated helium atom, which must be calculated consistently with
the final internal energy, $E$. An initial static precompression that
changes $V_0$ will also affect $E_0$ and $P_0$ but the corrections are
negligible as long as the amount of initial compression work as small
compared to the energy that is deposited dynamically. Assuming
$dE_0=dP_0=0$, the total differential of $H$ reads,
\beq
dH = dE \,+\, \frac{P}{2} dV \,-\, \frac{P}{2} dV_0 \,+\, \frac{1}{2} (V-V_0) dP
\label{hugdiff}
\eeq
The point of maximum compression, $\eta_{\rm max} = V_0/V$, along the
Hugoniot curve can be derived by setting, $dH=dV=dV_0=0$. The
resulting condition can be expressed in terms of the Gr\"uneisen
parameter, $\gamma \equiv V \left. \frac{\partial P}{\partial E}\right
|_V = 2/(\eta_{\rm max}-1)$.

Now we will determine how the maximum compression ratio, $\eta_{\rm
max}$, changes if the sample is precompressed statically. Keeping the
final shock pressure constant, the compression ratio changes as
function of the initial sample volume, $V_0$,
\beq
\left.\frac{\partial \eta}{\partial V_0}\right|_P = \frac{1}{V} - \frac{V_0}{V^2} \left. \frac{\partial V}{\partial V_0}\right|_P \;.
\label{hugeq2}
\eeq
Setting $dH=dP=0$ in Eq.~\ref{hugdiff}, one finds,
\beq
\left. \frac{\partial V_0}{\partial V} \right|_P = \left. \frac{2}{P} \frac{\partial E}{\partial V}\right|_P +1 \equiv \frac{2}{\delta} +1 \;.
\eeq
Inserting this result into Eq.~\ref{hugeq2} yields,
\beq
V \left.\frac{\partial \eta_{\rm max}}{\partial V_0}\right|_P = \frac{ 2(\gamma - \delta)}{\gamma (2+\delta)}\;.
\eeq
Since the parameters $\gamma$ and $\delta$ are both positive, the relation,
$\left. \frac{\partial \eta_{\rm max}}{\partial V_0} \right|_P > 0$, is equivalent
to the relation, $\delta < \gamma$, which is again equivalent to,
\beq
1 < \frac{\rho}{P} \left. \frac{\partial P}{\partial \rho}\right|_E \;.
\eeq
If this condition is fulfilled for a particular EOS then the maximum
shock compression ratio will decrease if the sample is precompressed
statically, which reduces $V_0$. We have computed the isoenergetic
compressibility for our first-principles EOSs for helium and hydrogen
and verified that this condition is satisfied for both materials
(Fig.~\ref{hug2}). It is also fulfilled for an ideal plasma model
because the maximum compression ratio is determined by the balance of
excitations of internal degrees of freedom and interaction
effects~\cite{Mi06}. Although all interactions are neglected, an ideal
model correctly represents the fact the excited states are suppressed
at high density because of the reduced entropy. The diminished
importance of excitations reduces the maximum compression ratio to
values closer to 4, which is the expected result for noninteracting
systems without internal degrees of freedom.

Recent laser shock wave experiments~\cite{eggert08} reached pressures
of 2 megabars in fluid helium for the first time. The sample was
precompressed statically in a modified diamond anvil cell before the
shock was launched. The static precompression is an important
development that enables one to reach higher densities and still
allows one to direct determine the EOS. Reaching higher densities is
important for planetary interiors because shock Hugoniot curves rise
faster than adiabats in a $P$-$T$ diagram shown in Fig.~\ref{hug1}. As
a result, a large part of Jupiter's adiabat remains inaccessible
unless one increases the starting density by precompression. The
precompression and relation of planetary interiors was studies
theoretically in Ref.~\cite{MH08}. It was demonstrated that
precompression of up to 60 GPa would be needed to characterize 50\% of
Jupiter's envelope. The challenge is here to reach high enough
densities because a single shock wave compresses the material only
5.25-fold or less (Fig.~\ref{hug2}).

The measurements of J. Eggert {\it et al.}~\cite{eggert08} confirmed
two of our theoretical predictions~\cite{Mi06}. They showed that
helium has a shock compression ratio substantially larger than 4 due
the electronic excitations and that the compression ratio would
decrease with increasing precompression (Fig.~\ref{hug2}). However,
the measurements appeared to be in better agreement with the SCvH EOS
model than PIMC simulations~\cite{eggert08}. The SCvH model predicts a
maximum compression ratio of 6.5 to occur around 300 GPa. A different
chemical model based on an expansion of the activity~\cite{ross07}
predicts maximum compression ratios between 5.6 and 6.2 to occur at
about 100 GPa.

Figure~\ref{hug3} shows a detailed comparison between experiments and
our first-principles simulations. The shock measurements without
precompression indeed show a higher compression than predicted from
first principles. The deviations are outside the experimental error
bars. However, this discrepancy goes away with increasing
precompression. The shocks with 3.4-fold precompression are in good
agreement with first-principles predictions. We have no explanation
for this trend at present.

The reason why the SCvH EOS yields larger compression ratios can be
understood by looking at the pressure that this model predicts. Using
our first-principles EOS, we derived the shock temperatures that
correspond the reported $P$-$\rho$ measurements. In the resulting
temperature range of 24$\,$000 -- 63$\,$000$\,$K, the SCvH EOS
significantly underestimates the pressure (see Fig.~\ref{PcompSC}),
which leads to higher predicted compressions
(Eq.~\ref{hug}). Furthermore, all measurements fall in the region
where the SCvH model relied on interpolation (Fig.~\ref{hug1}) and is
not expect be as reliable.

\section{Conclusions}

This paper combined path integral Monte Carlo and density functional
molecular dynamics simulation to derive one coherent equation of state
for fluid helium at high pressure and temperature. Helium is a
comparatively simple material since it does not form chemical bonds
nor has core electrons, but the our approach of combining two
simulation techniques can be generalized to study more complex
materials at extreme conditions. Certainly the presented approach to
fit the free energy and to derive adiabats works for any set of
EOS data points derived from first-principles simulations.

For the future, one might consider replacing DFT-MD with coupled
ion-electron Monte Carlo~\cite{delaney06}. However this is strictly a
groundstate method and one would still need to find a way to include
electronic excitations.

\acknowledgements

This material is based upon work supported by NASA under the grant
NNG05GH29G and by the NSF under the grant 0507321. We thank D. Saumon
for providing us with his He EOS table~\cite{SC95}, and acknowledge
receiving the preliminary manuscript~\cite{StixrudeJeanloz08} from
L. Stixrude and R. Jeanloz. We thank the authors of
ref.~\cite{eggert08} for sending us a table with their experimental
results.


\appendix

\section{Free energy spline interpolation}

We constructed the following 2D spline interpolation of the free
energy in order to reproduce the internal energy and pressures from
Tab.~\ref{EOS}. We use atomic units of Hartrees and Bohr radii. For
each density of $r_s=\{2.4, 2.0, 1.6, 1.2, 0.8\}$, we construct a
cubic spline $F_n(T)$. Table~\ref{splinetable} lists 16 knot points
$(T_i,F(T_i))$ for each density. In addition, the first derivate
$\frac{\partial F}{\partial T}$ are specified at the lowest and
highest temperatures. This is sufficient to construct a cubic spline
function $F(T)$~\cite{numerical_recipes}.

In a similar fashion, we derive a spline function that contains that
free energy derivative with respect to density, $\frac{\partial
F}{\partial n}(T)$, at the lowest and highest densities, $r_s=2.4$ and $0.8$
respectively.  $n$ is the density of the electrons, $n=N_e/V$. Those
knot points as well as the $T$ derivatives are included in
Tab.~\ref{splinetable} also. 

In order to obtain the free energy for a particular density and
temperature, $(n^*,T^*)$, we proceed as follows. First we evaluate the
spline functions $F(T^*)$ and $\frac{\partial F}{\partial n}(T^*)$ at
temperature $T^*$. Using these five knots points and density
derivatives, we construct a spline function, $F(\log(n))$. We use
$\log(n)$ as argument because it better represents the
high-temperature limit of weak interactions. Note that the constructed
splines for the density derivate contain $\frac{\partial F}{\partial
n}$ and not $\frac{\partial F}{\partial \log(n)}$. Then $F(\log(n))$
is evaluated at the density of interest, $n^*$. Finally we add the
term, $-T\Delta S = -13.7902836$ Ha*$T$, which brings the entropy in
agreement with our Debye-H\"uckel reference point at high temperature
for $r_s$=1.86.  This procedure yields the free energy $F(n^*,T^*)$ in
Hartrees per electron. Other thermodynamic variables including
pressure, internal energy, entropy, and Gibbs free energy can be
obtained by differentiation.

\begin{table*}
\caption{Knot points for free energy spline interpolation}
\begin{tabular}{cccccc}
\colrule
$T$(a.u.)  &$f(r_s=2.4,T)$& $f(r_s=2.0,T)$& $f(r_s=1.6,T)$&
$f(r_s=1.2,T)$ & $f(r_s=0.8,T)$\\
\colrule
0.001583407607 & -1.433567121 &   -1.431135811 &   -1.422237604 &   -1.377317056 &  -1.089747214 \\
0.004369348882 & -1.406243368 &	  -1.402248169 &   -1.390358674 &   -1.34121764 &   -1.049220122 \\
0.01205704051 &	 -1.338444742 &	  -1.330719999 &   -1.313137294 &   -1.255091313 &  -0.9491949603\\
0.03327091284 &	 -1.169814789 &	  -1.154154406 &   -1.125146478 &   -1.047701304 &  -0.7135001905\\
0.09180973061 &	 -0.7578793569 &  -0.7214892496 &  -0.6653645937 &  -0.5501514183 & -0.1583196636\\
0.2533452171 &	 0.1213432393 &	  0.2279596473 &   0.3697481593 &   0.5961317377 &  1.145222958  \\
0.6990958211 &	 1.448161918 &	  1.849742465 &	   2.331720062 &    2.967152398 &   4.011622282  \\
1.929126481 &	 1.650153009 &	  3.082886945 &	   4.795803813 &    6.956651264 &   9.947352284  \\
5.323346054 &	 -6.480107758 &	  -2.223161966 &   2.964697691 &    9.579180258 &   18.74636894  \\
14.6895569 &	 -50.37972679 &	  -38.41515844 &   -23.79211621 &   -4.992619324 &  21.39894481  \\
40.53523473 &	 -231.4313924 &	  -198.2806566 &   -157.6765535 &   -105.3657984 &  -31.61456362 \\
111.8553314 &	 -893.7314032 &	  -802.1449141 &   -689.7861073 &   -544.9061891 &  -340.7935545 \\
308.660237 &	 -3172.519712 &	  -2918.919455 &   -2608.696846 &   -2207.77895 &   -1644.63782  \\
851.7353686 &	 -10693.37579 &	  -9996.590015 &   -9137.681211 &   -8032.176108 &  -6477.239292 \\
2350.329104 &	 -34893.6373  &	  -32971.41578 &   -30600.55572 &   -27552.36594 &  -23259.81241 \\
6485.637557 &	 -111050.5942 &   -105746.9466 &   -99203.62441 &   -90788.1102  &  -78945.99564 \\
$f'(r_s,T_1)$   &    10.43646526  &   10.93007841  &  12.42958105 &  14.04097874 &  13.75171132  \\
$f'(r_s,T_N)$   &   -19.3620206   &  -18.54727965  & -17.53417624 & -16.23883073 & -14.41224513  \\
\colrule
\\
\end{tabular}

\begin{tabular}{cccccc}
\colrule
$T$(a.u.)      & $\frac{\partial f}{\partial n}(r_s=2.4,T)$ & $\frac{\partial f}{\partial n}(r_s=0.8,T)$\\
\colrule
0.001583407607 & 0.1605721538 & 0.9460950728 \\
0.004369348882 & 0.2923651353 & 0.9625433841 \\
0.01205704051 &	 0.6458549976 & 0.9782855175 \\
0.03327091284 &	 1.463747273 &  1.029890765 \\
0.09180973061 &	 3.574255133 &  1.142591093 \\
0.2533452171 &	 10.98479122 &  1.409424435 \\
0.6990958211 &	 43.27443042 &  2.089943658 \\
1.929126481 &	 153.211445 &   5.309130852 \\
5.323346054 &	 450.4159975 &  15.9843674  \\
14.6895569 &	 1263.720981 &  46.33079487 \\
40.53523473 &	 3502.359609 &  129.8984992 \\
111.8553314 &	 9679.619949 &  359.8256334 \\
308.660237 &	 26719.83852 &  992.5003633 \\
851.7353686 &	 73736.1465 &   2741.461026 \\
2350.329104 &	 203474.5198 &  7568.332998 \\
6485.637557 &	 561480.8894 &  20879.15481 \\
$f'(r_s,T_1)$ &   47.35562925 & 6.453623893 \\
$f'(r_s,T_N)$ &   86.57538487 & 3.218499127 \\
\colrule
\end{tabular}
\label{splinetable}
\end{table*}
             
\begin{table}
\label{coord}
\caption{Reduced coordinates of the DFT-MD configuration with 57 atoms that 
was used to report the $r_s=1.86$ results for the instantaneous pressure
in Fig.~\ref{Pinst}. The cell size is $L=14.5382$ a.u.}
$~x/L~~~~~~y/L~~~~~~~~z/L~~~$~~~$~~~x/L~~~~~~~y/L~~~~~~~z/L$\\
0.749029 0.334272 0.723992,~~
0.359050 0.631169 0.090795\\
0.636183 0.917961 0.531890,~~
0.500277 0.715818 0.420142\\
0.509121 0.642554 0.328933,~~
0.192192 0.222632 0.042651\\
0.273631 0.845722 0.363632,~~
0.070837 0.830223 0.693497\\
0.053785 0.837401 0.054990,~~
0.138489 0.091713 0.097622\\
0.250609 0.517490 0.740851,~~
0.953625 0.430789 0.067921\\
0.107008 0.407958 0.463387,~~
0.023708 0.960709 0.487179\\
0.988548 0.830572 0.241931,~~
0.811738 0.062550 0.902069\\
0.244399 0.482412 0.399190,~~
0.693258 0.647174 0.360832\\
0.924284 0.678572 0.470508,~~
0.181701 0.886709 0.333868\\
0.780287 0.033015 0.620919,~~
0.859185 0.932541 0.252564\\
0.774645 0.083064 0.349744,~~
0.903457 0.888628 0.124621\\
0.293881 0.081041 0.053630,~~
0.220134 0.760599 0.688370\\
0.493690 0.930407 0.343378,~~
0.585411 0.439278 0.167284\\
0.648043 0.965342 0.702852,~~
0.219455 0.957094 0.895428\\
0.504966 0.639074 0.084498,~~
0.906610 0.508304 0.938057\\
0.716468 0.854022 0.986517,~~
0.385839 0.307391 0.681601\\
0.099368 0.291429 0.740170,~~
0.475139 0.160612 0.598743\\
0.252564 0.696499 0.576596,~~
0.788211 0.564812 0.486616\\
0.613177 0.259980 0.238984,~~
0.296858 0.344416 0.229757\\
0.526564 0.816547 0.598836,~~
0.429733 0.712523 0.742929\\
0.507514 0.904602 0.268688,~~
0.685066 0.562001 0.926251\\
0.614731 0.263859 0.402947,~~
0.432246 0.210193 0.939664\\
0.115992 0.498747 0.676389,~~
0.424152 0.141821 0.676522\\
0.778767 0.981750 0.935757,~~
0.208696 0.768371 0.292528\\
0.334815 0.183086 0.275601,~~
0.487257 0.590889 0.227333\\
0.975542 0.456665 0.257836,~~
0.577884 0.835181 0.876629\\
0.737370 0.699890 0.544111,~~
0.177496 0.781162 0.853225\\
0.558513 0.066648 0.194491~~~~~~~~~~~~~~~~~~~~~~~~~~~~~~~~~~~~~~~~~\\
\end{table}

\end{document}